\documentclass[lettersize,journal]{IEEEtran}
\usepackage{amsmath,amsfonts}
\usepackage{algorithm}
\usepackage{array}
\usepackage[caption=false,font=normalsize,labelfont=sf,textfont=sf]{subfig}
\usepackage{textcomp}
\usepackage{stfloats}
\usepackage{float}
\usepackage{url}
\usepackage{verbatim}
\usepackage{graphicx}
\usepackage{bm}
\usepackage{array}
\usepackage{booktabs}
\usepackage{threeparttable}
\usepackage{multicol}  
\usepackage{multirow}  
\usepackage{amsthm, amssymb}
\usepackage{mathrsfs}
\usepackage[top=2cm, bottom=2cm, left=2cm, right=2cm]{geometry}
\usepackage{algorithmicx}
\usepackage{algpseudocode}
\usepackage{cite}
\usepackage{colortbl}
\usepackage[font=normalsize, textfont=sf, justification=centering]{caption}
\definecolor{mygray}{gray}{.9}
\floatname{algorithm}{Algorithm}
\renewcommand{\algorithmicrequire}{\textbf{Initialize:}}

\newcommand{\tabincell}[2]{\begin{tabular}{@{}#1@{}}#2\end{tabular}}  
\begin{document}
\title{{Min-Max Latency Optimization for IRS-aided Cell-Free Mobile Edge Computing Systems}}
\author{Nana Li, Wanming Hao, \emph{Member, IEEE}, Fuhui Zhou, \emph{Senior Member, IEEE}, Shouyi Yang, and Naofal Al-Dhahir,~\IEEEmembership{Fellow, IEEE}
	\thanks{The work  was supported in part by the National Natural Science Foundation of China under Grant 62101499. }
	\thanks{N. Li, W. Hao, and S. Yang are with the School of Information Engineering, Zhengzhou University, Zhengzhou 450001, China. (E-mail: nnli@gs.zzu.edu.cn, \{iewmhao, iesyyang\}@zzu.edu.cn)}
	\thanks{F. Zhou is with the College of Electronic and Information Engineering,
			Nanjing University of Aeronautics and Astronautics, Nanjing, 210000, China (E-mail: zhoufuhui@ieee.org).}
    \thanks{N. Al-Dhahir is with the Department of Electrical and Computer Engineering, University of Texas, Dallas, USA. (E-mail:  aldhahir@utdallas.edu)}
}

\maketitle

\begin{abstract}
Mobile-edge computing (MEC) is expected to provide low-latency computation service for wireless devices (WDs).  However, when WDs are located at cell edge or communication links between base stations (BSs) and WDs are blocked, the offloading latency will be large. To address this issue, we propose an intelligent reflecting surface (IRS)-assisted cell-free MEC system consisting of multiple BSs and IRSs for improving the transmission environment. Consequently, we formulate a min-max latency optimization problem by jointly designing multi-user detection (MUD) matrices, IRSs' reflecting beamforming vectors, WDs' transmit power and edge computing resource, subject to constraints on edge computing capability and IRSs phase shifts. To solve it, an alternating optimization algorithm based on the block coordinate descent (BCD) technique is proposed, in which the original non-convex problem is decoupled into two subproblems for alternately optimizing computing and communication parameters. In particular, we optimize the MUD matrix based on the second-order cone programming (SOCP) technique, and then develop two efficient algorithms to optimize  IRSs' reflecting vectors based on the semi-definite relaxation (SDR) and successive convex approximation (SCA) techniques, respectively. Numerical results show that employing IRSs in cell-free MEC systems outperforms conventional MEC systems, resulting in up to about $ 60 \% $ latency reduction can be attained. Moreover, numerical results confirm that our proposed algorithms enjoy a fast convergence, which is beneficial for practical implementation.
\end{abstract}


\section{Introduction}
Recently, the rapid development of the internet of thing (IoT) and artificial intelligence (AI) techniques have enabled various new applications (such as natural language processing, face/fingerprint recognition, autonomous driving, 3D media, etc.) based on real-time communication. Generally, these applications require low latency and need more computing resources.  However, this is challenging for IoT devices due to their limited computing capabilities \cite{ref1, ref2, ref3}. To address this challenge, mobile edge computing (MEC) was proposed, where the computing tasks of IoTs devices can be offloaded to the edge servers that are usually equipped with huge computing resources \cite{ref4, ref5, ref6, ref7}. However, the transmission latency needs to be considered in a MEC system, which is determined by the offloading links. When wireless devices (WDs) are located at the cell edge or communication links between base stations (BSs) and WDs are blocked, the offloading links will be poor, which leads to large transmission latency. Therefore, it is crucial to study how to improve the wireless communication environment for further exploiting the potential of MEC systems.

To provide better offloading links, the ultra-dense network (UDN) architecture with a lot of small BSs can be applied in order to shorten the distances between BSs and WDs or provide direct links between them \cite{ref8}. However, as the number of BSs increases, inter-cell interference becomes the bottleneck \cite{ref9, ref10, ref11, ref12}. Thus, the \emph{user-centric}-based cell-free network structure is proposed, where BSs simultaneously serve all WDs to avoid the multi-cell interference. On the other hand, the energy consumption and hardware cost are still high for the cell-free network due to the deployment of massive BSs. Recently, an intelligent reflecting surface (IRS) composed of a large number of low-power passive reflecting elements was developed, which can focus signal energy in the desired spatial direction by adjusting phase shifts, thus enlarging the wireless converge \cite{ref13, ref14, ref15, ref16}. Therefore, to reduce the system cost, some BSs can be replaced by IRSs. In this paper, we will investigate the latency optimization problem in the IRS-aided cell-free MEC systems.

\subsection{Related Works}
Recenty, there have been many works focusing on MEC. Initially, the binary offloading model was considered, and each computation task is treated as a whole. Specifically, Zhang et al. \cite{ref17} investigated to minimize the energy consumption (EC) of a single user by dynamically configuring the clock frequency of the local central processing unit (CPU) for local computing and varying the offloading rate for cloud computing according to the stochastic channel condition. Then, Wang et al. \cite{ref18} extended the single-user MEC system to the multi-user one, and aimed to minimize the weighted sum of all users' EC. For the binary offloading model, the offloading design in multi-user MEC is usually a typical mixed-integer nonlinear programming problem that is usually nonconvex. To deal with it, the authors in \cite{ref18}  proposed a low-complexity two-stage algorithm by iteratively optimizing the offloading decision and resource allocation. The authors in \cite{ref19, ref20, ref21} formulated the offloading decision optimization problem as a multi-user noncooperative computing offloading game, and then an effective algorithm based on game theory was proposed. The authors in \cite{ref24, ref25} considered a partical offloading model, in which a task can be further divided into multiple modules, which can be performed locally or remotely. The weighted sum of EC and computing latency was minimized by jointly optimizing offloading decision and CPU frequency in \cite{ref24}, or offloading desicion and transmit power as in \cite{ref25}. Additionally, several researchers also investigated the edge computing problem under multi-tier computing systems.  For example, the authors in \cite{ref22, ref23, ref44} integrated cloud computing with MEC and proposed a three-tier system model ( “terminal-edge-cloud” ). Particularly, Ning et al. \cite{ref22} considered both single-user and multi-user scenarios, and for the single-user scenario, a branch and bound algorithm was proposed. For the multi-user scenario, an iterative heuristic MEC resource allocation algorithm was developed for making the offloading decision dynamically. In \cite{ref23}, the optimal offloading node selection problem was formulated as a Markov decision process, and then an iterative optimization algorithm was proposed. Gao et al. \cite{ref44} considered a three-tier system consisting of multiple WDs, multiple edge nodes, and a central cloud, and then proposed two offloading algorithms by iteratively optimizing the WD-edge matching strategy and resource allocation.

Additionally, there are several works investigating edge computing problem in IRS-aided MEC systems.  Haber et al. \cite{ref26} proposed an IRS-aided single-user MEC system, where the computing task can be offloaded to multiple MEC nodes for guaranteeing high reliability. Later, Sun et al. \cite{ref27} considered both the flat-fading channel and frequency-selective channel model under the IRS-aided multi-user MEC scenario, and proposed an alternating optimization algorithm to minimize users' energy consumption. The authors in \cite{ref28, ref29} investigated the energy consumption problem, where,  compared with the conventional MEC system, about 80\%  energy consumption reduction was achieved in the proposed IRS-aided wireless powered MEC system. Besides, to avoid the interference among WDs, the authors in \cite{ref30, ref31} applied the time division multiple access protocol in the computing offloading stage. Similarly, the authors of \cite{ref30} also considered the IRS-aided wireless powered MEC system, and they jointly optimized downlink/uplink IRS reflecting beamforming vector, transmit power, time allocation for energy transmission of downlink and  computing offloading of uplink, and local CPU frequency. In \cite{ref31}, the authors jointly optimized offloading time, CPU frequency, transmit power, and IRS phase shift vector in the IRS-aided MEC system. Note that both \cite{ref30} and \cite{ref31} aim to maximize the total computation data bits of all users. To minimize the total latency, Zhou et al. \cite{ref32} proposed a new flexible \emph{time-sharing} non-orthogonal multiple-access (NOMA) scheme, in which the offloading data can be divided into two parts, and then the optimal solution was obtained for both the cases of finite and infinite edge computation capacities. 

Based on the above brief literature survey, we note that, although there have been several works considering the latency problem in IRS-aided MEC system, WDs' fairness was not investigated. Furthermore, for future IRS-aided cell-free MEC systems, how to jointly optimize multiple BSs' received matrix, multiple IRSs' reflecting beamforming, WDs' transmit power and edge computing resource based on WDs' fairness is challenging. 

\subsection{Contributions}
In this paper, considering WDs' fairness during the computation offloading, we investigate the min-max WD's latency in IRS-aided cell-free MEC systems, and the main contributions are summarized as follows.
\begin{itemize}
	\item[•] We investigate the latency minimization problem in the designed IRS-aided cell-free MEC systems. To guarantee the WDs' fairness, we formulate a min-max WD's latency optimization problem by jointly designing offloading date size, edge computing resource, multi-user detection (MUD) matrix and reflecting beamforming vector. Meanwhile, the limited edge computing resource is considered. Since it is challenging to directly solve the formulated optimization problem, the block coordinate descent (BCD) technique is adopted to decouple it into two subproblems based on the alternate optimization approach.
	
	\item[•] The first subproblem is to jointly optimize the offloading data size and edge computing resource based fixed MUD matrix and reflecting beamforming vector. To solve it, we first analyze the relation between offloading data size and edge computing resource, and then transform it into a convex optimization problem with the aid of the bisection search method. Next, an iterative algorithm is proposed to solve the first subproblem. 
	
    \item[•] The second subproblem is to jointly optimize the MUD matrix and reflecting beamforming vector based on obtaining offloading data size and edge computing resource from the first subproblem. To solve it, we propose an alternate optimization iterative algorithm based on MUD matrix and reflecting beamforming vector. Then, we solve for the MUD matrix based on the second-order cone programming (SOCP) technique, and solve for the reflecting beamforming vector based on the semi-definite relaxation (SDR) and successive convex approximation (SCA) techniques. Finally, two subproblems are alternately and iteratively solved until convergence. Furthermore, we analyze the convergence of the proposed schemes. 
\end{itemize} 

\begin{table}[t]
	\begin{center}
		\caption{Notation List. }
		\label{tab1}
		\begin{tabular}{ l | r }
			\hline
			Parameter &  Definition\\
			\hline
			$ \left(\cdot \right)^{H} $       & Conjugate transpose  \\
			\hline
			$ \mathbb{E}\left\lbrace \cdot \right\rbrace    $       &  Expectation  \\
			\hline
			$ \left(\cdot \right)^{T} $   & Transpose   \\
			\hline
			$ \operatorname{Tr} \left\lbrace \cdot \right\rbrace $        & Trace  \\
			\hline
			$\operatorname{arg}  \left\lbrace \cdot \right\rbrace $  & The argument of a complex number\\
			\hline
			$ \mathbb{C}^{m\times n} $     &  The space of $ m\times n $ complex-valued matrices \\
			\hline
			$ \operatorname{det} \left(\mathbf{A} \right) $       &  Determinant of $ \mathbf{A} $  \\
			\hline
			$ \operatorname{diag} \left\lbrace \cdot \right\rbrace $ & Diagonalization operator \\
			
			\hline
			$ \left| \, \cdot \, \right| $   &  Absolute value of a scalar \\
			\hline
			$ \left\| \, \cdot \, \right\|_{2} $   &   Euclidean norm / 2-norm  \\    
			\hline
			$ \mathbf{V} \succeq 0 $   & Positive semi-definite matrix \\
			\hline
			$ \operatorname{Re} $   &     Real part   \\
			\hline
			$ \operatorname{Im} $   &   Imaginary part  \\
			\hline
			$ \left\lfloor.\right\rfloor $  & The floor operations \\
			\hline   
			$ \left\lceil.\right\rceil $  & The ceiling operations \\
			\hline
			$ \mathbf{V}_{m,n} $        & The element of $ \mathbf{V} $ in the $ m $-th row, $ n $-th row. \\ 
			\hline 
			$ L_{k} $                  &  The total input data size\\
			\hline 
			$ \ell_{k} $               &  The offloading volume \\
			\hline
			$ c_{k} $                  &  The computation complexity \\
			\hline                           
			$ f_{k}^{l} $              &   The local computing capability of the $ k $-th WD \\
			\hline				                         
			$ f_{k}^{e} $             &  \tabincell{r}{The edge computing resource\\ allocated to the $ k $-th WD } \\
			\hline	                           
			$ D_{k}^{l} $            &   \tabincell{r}{The time required for carrying \\ out the local computation } \\
			\hline 
			$ D_{k}^{e} $            &  \tabincell{r}{ The total latency induced by the \\ data offloading and edge computing} \\
			\hline                            
		\end{tabular}
	\end{center}
\end{table}

The rest of this paper is organized as follows. Section II introduces the system model and the formulated problem. In Sections III, we develop an alternating algorithm for solving the formulated min-max latency optimization problem. Section IV investigates the latency optimization problem in an IRS-aided cell-free MEC system with single WD. Section V provides the performance evaluation. Finally, the conclusions are given in Section VI. For clarity, we list the mathematical operations adopted throughout this paper in Table I. 
\begin{figure*}[t]
	\begin{minipage}[t]{0.42\textwidth}
		\centering
		\includegraphics[width=1.2\textwidth]{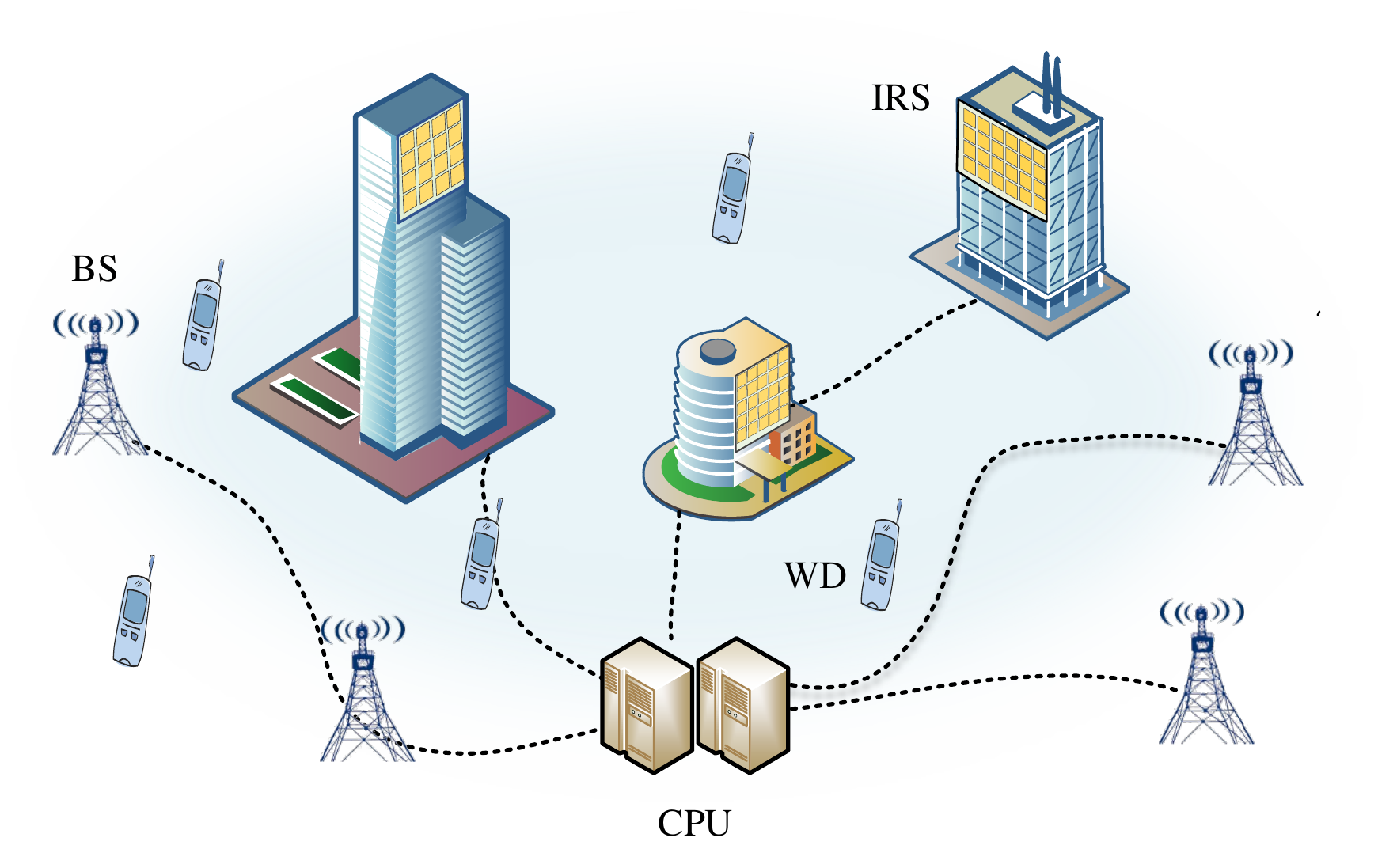}
		\caption{The proposed IRS-aided cell-free MEC system.}
	\end{minipage}
	\qquad \qquad
	\begin{minipage}[t]{0.42\textwidth}
		\centering
		\includegraphics[width=1.2\textwidth]{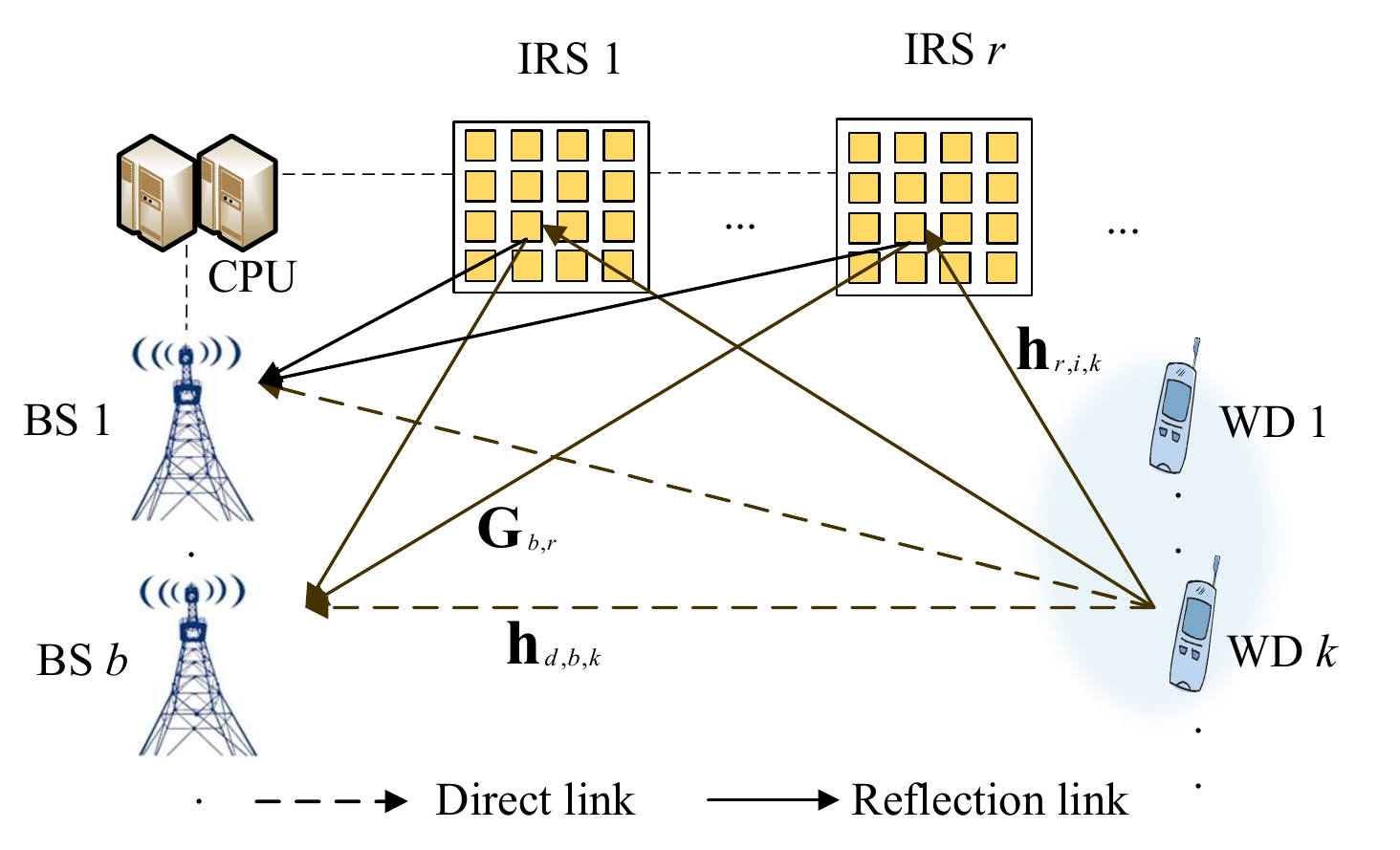}
		\caption{The channels model in the IRS-aided cell-free MEC system.}
	\end{minipage}
\end{figure*}
\section{System Model and Problem Formulation}
\subsection{Communication Model}
As shown in Fig. 1, we consider an IRS-aided cell-free MEC system, where multiple distributed BSs cooperatively serve WDs with the help of IRSs. All IRSs and BSs are connected to a central processing unit by high-throughput optical cables. The considered system consists of $ K $ single-antenna WDs, $ I $ IRSs, and $ B $ BSs. The number of elements at the $ i $-th IRS and number of antennas at the $ b $-th BS are denoted by  $ N_{i} $ and $ M_{b} $, respectively. For simplicity, we set $ N_{i}=N $ and  $ M_{b}=M $ for any $ i  $ and $ b $,  and let  $ \mathcal{N} \in\{1,\dots, N\} $, $ \mathcal{B} \in\{1,\dots, B\}  $, $ \mathcal{I} \in\{1,\dots, I\} $, and $ \mathcal{K} \in\{1,\dots, K\} $ denote the index sets of IRS elements, BSs, IRSs, and WDs, respectively. We consider a block-fading channel model, where the wireless channels remain constant during the current time block but change over different time blocks,  and assume that the channel state information (CSI) of all involved channels are available by using existing advanced channel estimation methods \cite{ref33, ref34}.

The direct channel vector from the $ k $-th WD to the $ b $-th BS is denoted by $ \mathbf{h}_{d,b,k} \in \mathbb{C}^{M\times 1} $, and the reflect channel vectors from the $ k $-th WD to the $ i $-th IRS and the $ i $-th IRS to the $ b $-th BS are denoted by $ \mathbf{h}_{r,i,k} \in \mathbb{C}^{N\times 1} $ and  $ \mathbf{G}_{b,i} \in \mathbb{C}^{M\times N} $, respectively. The phase shift coefficient vector of the $ i $-th IRS is denoted by $\boldsymbol{\theta}_{i}=\left[\theta_{i,1}, \theta_{i,2}, \ldots, \theta_{i,N}\right]^{T}$, where $ \theta_{i,n} \in \left[ 0, 2 \pi \right) $ for all $ i \in \mathcal{I}  $ and $ n \in \mathcal{N}  $. Then, the diagonal reflecting matrix of the $ i $-th IRS is given by
 \begin{equation} 
 	\mathbf{\Theta}_{i} \triangleq {\rm diag}\left( \beta_{i,1} e^{j \theta_{i,1}},..., \beta_{i,N}e^{j \theta_{i,N}} \right), \forall i\in \mathcal{I}, 
 \end{equation}
where $ \beta_{i,n} \in \left[ 0,1 \right] $ stands for the reflection amplitudes of the IRS elements and each of them is fixed to one for maximizing the reflected signal power. Thus, the combined effective channel from the $ k $-th WD to the $ b $-th BS can be defined as
\begin{eqnarray}
\mathbf{h}_{b,k}=\mathbf{h}_{d,b,k}+\sum_{i=1}^{I} \mathbf{G}_{b,i} \mathbf{\Theta}_{i} \mathbf{h}_{r,i,k}.
\end{eqnarray}

Let $ P_{t} $ and $ \mathbf{s}=\left[ s_{1}, s_{2},...,s_{K} \right]^{T}  $ denote the offloading power and signal of $ K $ WDs, respectively. Here, we assume that all WDs transmit the same power for simplicity as in \cite{ref42}. Let $ \mathbf{w}_{b,k} \in \mathbb{C}^{M \times 1} $ denote the MUD vector of the $ b $-th BS for the $ k $-th WD.  Hence, the detected signal at BSs for the $ k $-th WD can be formulated as
\begin{small}
\begin{align}
	&\hat{s}_{k}=\sum_{b=1}^{B}\mathbf{w}_{b,k}^{H}\left[\sqrt{P_{t}} \sum_{j=1}^{K} \left(\mathbf{h}_{d,b,j}\!+\!\sum_{i=1}^{I} \mathbf{G}_{b,i} \mathbf{\Theta}_{i} \mathbf{h}_{r,i,j}\right)s_{j}\! +\!\mathbf{n}_{b} \right]  \nonumber \\
	&\;\;\;\overset{\left( a \right) }{=}\mathbf{w}_{k}^{H}\left[\sqrt{P_{t}} \sum_{j=1}^{K} \left(\mathbf{h}_{d,j}\!+\!\sum_{i=1}^{I} \mathbf{G}_{i} \mathbf{\Theta}_{i} \mathbf{h}_{r,i,j}\right)s_{j} +\mathbf{n} \right] \nonumber \\
	&\;\;\;\overset{\left( b \right) }{=}\mathbf{w}_{k}^{H}\left[\sqrt{P_{t}} \sum_{j=1}^{K} \left(\mathbf{h}_{d,j}\!+\! \mathbf{G} \mathbf{\Theta} \mathbf{h}_{r,j}\right)s_{j} +\mathbf{n} \right] \\
	&\;\;\;\overset{\left( c \right) }{=}\mathbf{w}_{k}^{H}\left[\sqrt{P_{t}} \sum_{j=1}^{K} \mathbf{h}_{j}s_{j} +\mathbf{n} \right], \nonumber
\end{align}	
\end{small} 
\!\!\!where $ \mathbf{n}_{b} \in \mathbb{C}^{M \times 1} $ denotes the received noise vector of the $ b $-th BS, $ \mathbf{w}_{k} \in \mathbb{C}^{MB \times 1} $ is the $ k $-th column of the MUD matrix $ \mathbf{W} \in \mathbb{C}^{MB \times K} $, where $ \mathbf{W}=\left[ \mathbf{w}_{1}^{T}, ..., \mathbf{w}_{K}^{T}\right]^{T}  $,  $ \left( a \right) $ holds by defining $ \mathbf{w}_{k}=\left[\mathbf{w}_{1,k}^{T},..., \mathbf{w}_{B,k}^{T} \right]^{T}  $, $ \mathbf{h}_{d,k}=\left[ \mathbf{h}_{d,1,k}^{T},...,\mathbf{h}_{d,B,k}^{T} \right]^{T} $, $ \mathbf{G}_{i}=\left[\mathbf{G}_{1,i}^{T},...,\mathbf{G}_{B,i}^{T} \right]^{T}  $, and $ \mathbf{n}=\left[\mathbf{n}_{1}^{T},...,\mathbf{n}_{B}^{T} \right]^{T}  $, $ \left( b \right) $ holds by defining $ \mathbf{G}=\left[\mathbf{G}_{1}^{T},...,\mathbf{G}_{I}^{T} \right]^{T}  $,  $ \mathbf{\Theta}={\rm diag} \left(  \mathbf{\Theta}_{1},...,\mathbf{\Theta}_{I}\right)   $, and $ \mathbf{h}_{r,k}=\left[\mathbf{h}_{r,1,k}^{T},...,\mathbf{h}_{r,I,k}^{T} \right]^{T}$, and $ \left( c \right) $ holds according to $ \mathbf{h}_{k}=\mathbf{h}_{d,k}+ \mathbf{G} \mathbf{\Theta} \mathbf{h}_{r,k}$. Thus, the  received SINR for the $ k $-th WD is
\begin{equation}\label{Eq02}
	\gamma_{k}\!\left(\mathbf{w}_{k}, \boldsymbol{\theta} \right)\!=\!\dfrac{P_{t}\left| \mathbf{w}_{k}^{H}\left(\mathbf{h}_{d,k}\!+\! \mathbf{G} \mathbf{\Theta} \mathbf{h}_{r,k}\right) \right|^{2} }{P_{t}\!\sum_{j=1,j\ne k}^{K} \! \left| \mathbf{w}_{k}^{H}\!\left(\mathbf{h}_{d,j}\!+\! \mathbf{G} \mathbf{\Theta} \mathbf{h}_{r,j}\right) \right|^{2}\!\!+\!\sigma^{2} }, \\
\end{equation}
and the corresponding achievable rate is given by
\begin{eqnarray}\label{Eq01}
	R_{k}\left( \mathbf{w}_{k}, \theta\right)=B\log_2\left(1+\gamma_{k}\left(\mathbf{w}_{k}, \boldsymbol{\theta} \right)\right). 
\end{eqnarray}

\subsection{Computing Model}
In this paper, the partial offloading scheme is considered, and the computing models for both local and edge processing are presented, respectively.
\begin{itemize}
	\item \emph{Local computing:} Let $ f_{k}^{l} $, $ L_{k} $, $ \ell_{k} $ and $ c_{k} $ denote the CPU-cycle frequency (cycles/s), total computing data size, offloading data size, and computational complexity of input data for the $ k $-th WD, respectively. Thus, the latency imposed by local computing is formulated as $ D_{k}^{l}\left( \ell_{k}\right)=\left(L_{k}-\ell_{k} \right)c_{k}/f_{k}^{l}  $.
	\item \emph{Edge computing:} The latency for edge computing usually includes three parts: a) offloading latency for transmitting computing data to BSs; b) processing latency for executing offloaded data at MEC server; c) return latency for transmitting computing results to WDs. Let  $ f_{total}^{e} $ and $ f_{k}^{e} $ denote the total computing resource of MEC server and the computing resource allocated to the $ k $-th WD, respectively, satisfying  $ \sum_{k=1}^{K}f_{k}^{e} \leq f_{total}^{e} $. Here, we ignore the return latency since the returning results are usually of small size \cite{ref18,ref35}. Therefore, the total latency for edge processing can be written as  $D_{k}^{e}\left(\mathbf{w}_{k}, \boldsymbol{\theta}, \ell_{k}, f_{k}^{e}\right)=\ell_{k} / R_{k}\left(\mathbf{w}_{k}, \boldsymbol{\theta}\right)+\ell_{k} c_{k} / f_{k}^{e}$.
\end{itemize}
On this basis, the overall latency for the $ k $-th WD can be formulated as:
\begin{small}
\begin{equation}\label{Eq6}
	\begin{aligned}
		D_{k}\left(\mathbf{w}_{k}, \boldsymbol{\theta}, \ell_{k}, f_{k}^{e}\right) &=\max \left\{D_{k}^{l}\left(\ell_{k}\right), D_{k}^{e}\left(\mathbf{w}_{k}, \boldsymbol{\theta}, \ell_{k}, f_{k}^{e}\right)\right\} \\
		&=\max \left\{\frac{\left(L_{k}-\ell_{k}\right) c_{k}}{f_{k}^{l}}, \frac{\ell_{k}}{R_{k}\left(\mathbf{w}_{k}, \boldsymbol{\theta}\right)}+\frac{\ell_{k} c_{k}}{f_{k}^{e}}\right\}.
	\end{aligned}
\end{equation}
\end{small}

\subsection{Problem Formulation}
Considering WDs' fairness, we minimize the maximum WD's latency by jointly optimizing offloading data size $ \boldsymbol{\ell}=\left[ \ell_{1}, \ell_{2},...,\ell_{K}\right]^{T}  $, edge computing resource  $ \mathbf{f}^{e}= \left[ f_{1}^{e},f_{2}^{e},...,f_{K}^{e} \right]^{T} $, MUD matrix $ \mathbf{W} $, and reflecting beamforming vector $ \boldsymbol{\theta} $, which can be formulated as 
\begin{subequations} \label{OptA}
	\begin{align}
		\mathcal{P} 0: &\underset{ \boldsymbol{\ell}, \mathbf{f}^{e}\!,\mathbf{W}, \boldsymbol{\theta}}{\min }  \underset{k \in \mathcal{K} } {\max} ~  D_{k}\left(\mathbf{w}_{k}, \boldsymbol{\theta}, \ell_{k}, f_{k}^{e}\right) \notag \\
		\text { s.t. } & 0 \leq \theta_{i,n}<2 \pi, \quad \forall i\in \mathcal{I}, \forall n\in \mathcal{N}, \label{OptA1} \\
		& \ell_{k} \in\left\{0,1, \ldots, L_{k}\right\}, \forall k\in \mathcal{K},  \label{OptA2}\\
		& \sum_{k=1}^{K} f_{k}^{e} \leq f_{\text {total }}^{e}, \label{OptA3}\\
		& f_{k}^{e} \geq 0, \quad \forall k\in \mathcal{K} ,\label{OptA4} \\
		& \left\| \mathbf{w}_{k} \right\|^{2} \leq 1,  \quad  \forall k\in \mathcal{K},\label{OptA5}
	\end{align}
\end{subequations}
where (\ref{OptA1}) is IRS reflection coefficients constrains, (\ref{OptA2}) implies that the $ k $-th WD's offloaded data size should be an integer between zero and the total input data $ L_{k} $. (\ref{OptA3}) denotes the computing resource allocated to all WDs should not exceed the total edge computing resource. (\ref{OptA5}) represents unit-norm detection vector constrains for the $ k $-th WD. It is clear that $ \mathcal{P} $0 is difficult to be directly solved. Next, we first introduce an auxiliary variable $ t $ to transform $ \mathcal{P} 0 $ into the following $ \mathcal{P} 1 $, given as 
\begin{subequations} \label{OptB}
	\begin{align}
		\mathcal{P} 1: & \min _{\boldsymbol{\ell}, \mathbf{f}^{e}, \mathbf{W}, \boldsymbol{\theta}} t \notag  \\
	\text { s.t. } & D_{k}\left(\ell_{k}, f_{k}^{e}\right) \leq t, \forall k\in \mathcal{K},\label{OptB1} \\
		& (\ref{OptA1}), (\ref{OptA2}),(\ref{OptA3}),(\ref{OptA4}),(\ref{OptA5}). \label{OptB2}  
	\end{align}
\end{subequations}

\emph{ Remark 1: Although the objective function (OF) of $ \mathcal{P} 1 $ and constraints in ($ \ref{OptB2} $) are linear, it is still challenging to directly solve $ \mathcal{P} 1 $ due to the following three aspects: a) the segmented form of (\ref{OptB1}), b) MUD matrix $ \mathbf{W} $ and reflecting beamforming vector $ \boldsymbol{\theta} $ are coupled  together, c) (\ref{OptB1}) is non-convex with respect to $ \boldsymbol{\theta} $. In general, there is no standard method to find a globally optimal solution of such a non-convex optimization problem. To proceed, we develop an iterative framework to obtain a locally optimal solution. Specifically, the segmented form of (\ref{OptB1}) is reformulated as a linear form relying on the BCD technique. Then, fixing the computing setting, we optimize the MUD matrix and reflecting beamforming vector alternately. Finally, we develop two efficient algorithms based on SDR and SCA techniques to obtain a locally optimal solution of $ \boldsymbol{\theta} $, respectively.}

\section{Problem Solutions } 
In this section, we first divide $ \mathcal{P} 1 $ into two independent subproblems  relying on the BCD technique. Specifically, given $ \mathbf{W} $ and $ \boldsymbol{\theta} $, the offloading data size $ \boldsymbol{\ell} $ and edge computing resource $ \mathbf{f}^{e} $ are optimized. Then, based on the obtained $ \boldsymbol{\ell} $ and $ \mathbf{f}^{e} $, the MUD matrix $ \mathbf{W} $ and reflecting beamforming vector $ \boldsymbol{\theta} $ are optimized. The above procedure is repeated until convergence. 

\subsection{Jointly Optimizing $ \boldsymbol{\ell} $ and $ \mathbf{f}^{e} $ for Given $ \mathbf{W} $ and $ \boldsymbol{\theta} $} 
For given $ \mathbf{W} $ and $ \boldsymbol{\theta} $,  $ \mathcal{P} $1 can be reformulated as follows
\begin{subequations} \label{OptC}
	\begin{align}
		\mathcal{P}2: & \min _{\boldsymbol{\ell}, \mathbf{f}^{e},t} t \notag  \\
	   \text { s.t. }  & D_{k}\left(\ell_{k}, f_{k}^{e}\right) \leq t, \forall k\in \mathcal{K}, \label{OptC1}\\
	    & \ell_{k} \in\left\{0,1, \ldots, L_{k}\right\}, \forall k\in \mathcal{K}, \label{OptC2} \\
	    & \sum_{k=1}^{K} f_{k}^{e} \leq f_{\text {total }}^{e}, \label{OptC3} \\
	    & f_{k}^{e} \geq 0, \quad \forall k\in \mathcal{K}.  \label{OptC4}
	\end{align}
\end{subequations}
Based on the following Proposition 1, we optimize the offloading data size $ \boldsymbol{\ell} $. 

\emph{Proposition 1: For given $ \mathbf{f}^{e} $, the optimal offloading data size is given by
	\begin{equation} \label{Eq10}
		\ell_{k}^{*}=\underset{\hat{\ell}_{k} \in\left\{\left\lfloor\hat{\ell}_{k}^{*}\right\rfloor,\left\lceil \hat{\ell}_{k}^{*}\right\rceil\right\}}{\arg \min } D_{k}\left(\hat{\ell}_{k}\right),
	\end{equation}
where $ \left\lfloor.\right\rfloor $ and $ \left\lceil . \right\rceil $ denote the floor and ceiling operations, respectively, and the value of $ \hat{\ell}_{k}^{*} $ is selected for ensuring that $ D_{k}^{l} ( \hat{\ell}_{k} ) =  D_{k}^{e} ( \hat{\ell}_{k} ) $, i.e.,
\begin{equation}\label{Eq11}
	\hat{\ell}_{k}^{*}=\frac{L_{k} c_{k} R_{k} f_{k}^{e}}{f_{k}^{e} f_{k}^{l}+c_{k} R_{k}\left(f_{k}^{e}+f_{k}^{l}\right)}.
\end{equation}
}
\emph{Proof:} See Appendix A.
 
After obtaining the relation between offloading data size and edge computing resource, we substitute (\ref{Eq11}) into (\ref{OptC1}), then $ \mathcal{P}2 $ can be written as:
\begin{subequations} \label{OptD}
	\begin{align}
		\mathcal{P} 2.1: & \min _{\mathbf{f}^{e},t} t \notag  \\
		 \text { s.t. } &\frac{\left(L_{k} c_{k}^{2} R_{k}+L_{k} c_{k} f_{k}^{e}\right)}{f_{k}^{e}f_{k}^{l}+c_{k} R_{k}\left(f_{k}^{e}+f_{k}^{l}\right)} \leq t,  \forall k\in \mathcal{K}, \label{OptD1} \\
		 & \sum_{k=1}^{K} f_{k}^{e} \leq f_{\text {total }}^{e}, \label{OptD2} \\
		 & f_{k}^{e} \geq 0, \quad \forall k\in \mathcal{K}.  \label{OptD3}
	\end{align}
\end{subequations}
Note that $ \mathcal{P} 2.1 $ is a non-convex problem. To proceed it, we first reformulate (\ref{OptD1}) as $ \left(L_{k} c_{k}^{2} R_{k}+L_{k} c_{k} f_{k}^{e}\right) - t\left( f_{k}^{e}+c_{k} R_{k}\left(f_{k}^{e}+f_{k}^{l}\right)\right) \leq 0 $. Then, with the help of a bisection search over $ t $, $ \mathcal{P} 2.1 $ can be equivalently transformed into the following feasibility problem, given as 
\begin{subequations} \label{OptL}
	\begin{align}
		 \mathcal{P} 2.2: & {\rm Find: } ~\mathbf{f}^{e} \notag  \\
		 \text { s.t. } & \left(L_{k} c_{k}^{2} R_{k}+L_{k} c_{k} f_{k}^{e}\right) \notag\\
		&~~ - t^{l_{1}}\left( f_{k}^{e}+c_{k} R_{k}\left(f_{k}^{e}+f_{k}^{l}\right)\right) \leq 0,  \forall k\in \mathcal{K}, \label{OptL1} \\
		& \sum_{k=1}^{K} f_{k}^{e} \leq f_{\text {total }}^{e}, \label{OptL2} \\
		& f_{k}^{e} \geq 0, \quad \forall k\in \mathcal{K},  \label{OptL3}
	\end{align}
\end{subequations}
where $ t^{l_{1}} $ is the value of $ t $ at the $ l_{1} $-th iteration. For a given latency target $ t^{l_{1}} $, the above optimization problem is convex, and thus the global optimal solution of $ \mathcal{P} 2.2 $ can be found via existing convex optimization techniques, e.g., CVX solver. 

Note that if  $ \mathcal{P} 2.2 $ is feasible, the given latency $ t $ can be achieved. Assuming that the optimal solution of $ \mathcal{P} 2.1 $ is $ \hat{t} $, we can infer that, for any given $ t $, if $ \mathcal{P} 2.2 $ is feasible, we have $ t \geq \hat{t} $, while if $ \mathcal{P} 2.2 $ is infeasible, we have $ t \leq \hat{t} $. Therefore, with the aid of a bisection search over $ t $, $ \mathcal{P} 2.1 $ can be equivalently solved by checking the feasibility of $ \mathcal{P} 2.2 $ for a given $ t \geq 0 $.

%

In summary, the procedure for solving $ \mathcal{P} 2 $ is presented as \textbf{Algorithm 1}. 
\renewcommand{\algorithmicrequire}{\textbf{Input:}}
\begin{algorithm}[t]
	\caption{ Joint optimization scheme for solving $ \mathcal{P} 2 $ }
	\begin{algorithmic}[0] 
		\Require 
		$ \mathbf{h}_{k} $, $ B $, $ p_{t} $, $ \sigma^{2} $, $ L_{k} $, $ c_{k} $, $ K $, $ f^{e}_{total} $, $ \epsilon_{1} $, $ \mathbf{W} $, and $ \boldsymbol{\theta $}
		\Ensure
		Optimal $ \boldsymbol{\ell} $ and $ \mathbf{f}^{e} $ 
		\State \textbf{1. Initialization}\\
		Initialize $ l_{1}=0 $, calculate $ R_{k} $ according to ($ \ref{Eq01} $), $ \forall k\in \mathcal{K} $  
		\State \textbf{2. Joint optimization of $ t $ and $ \mathbf{f}^{e} $ }
		\Repeat:
		\State Calculate $ t^{\left( l_{1}\right) } $ using the bisection search method
		\State Solving $ \mathcal{P} 2.2 $ to obtain optimal $ \mathbf{f}^{e \left( l_{1}\right)} $
		\State  Update $ l_{1} \gets l_{1}+1 $
		\Until{ Convergence }
		\State \textbf{3. Calculate $ {\ell}_{k} $ using (\ref{Eq11}), $ \forall k\in \mathcal{K} $ } 
	\end{algorithmic}
\end{algorithm}

\subsection{Jointly Optimizing $ \mathbf{W} $ and $ \boldsymbol{\theta} $ for Given $ \boldsymbol{\ell} $ and $ \mathbf{f}^{e} $} 
Based on obtaining $ \boldsymbol{\ell} $ and $ \mathbf{f}^{e} $, $ \mathcal{P} $1 can be simplified as
\begin{subequations} \label{OptE}
	\begin{align}
		\mathcal{P} 3: & \min _{\mathbf{W}, \boldsymbol{\theta}, t} t \notag  \\
		\text { s.t. } & \dfrac{\ell_{k}}{R_{k}\left(\mathbf{w}_{k},\boldsymbol{\theta} \right) }+\dfrac{\ell_{k} c_{k}}{f_{k}^{e}} \leq t, \quad  \forall k\in \mathcal{K} \label{OptE1} \\
		& 0 \leq \theta_{i,n}<2 \pi, \quad \forall i\in \mathcal{I}, \forall n\in \mathcal{N}, \label{OptE2}\\
		& \left\| \mathbf{w}_{k} \right\|^{2} \leq 1,  \quad  \forall k\in \mathcal{K} \label{OptE3}.   
	\end{align}
\end{subequations}

\emph{ Remark 2: Solving $ \mathcal{P} 3 $ involves a joint optimization of MUD matrix $ \mathbf{W} $, reflecting beamforming vector {$ \boldsymbol{\theta} $} and latency $ t $,  which is  challenging. Next, we develop an alternately iterative optimization algorithm to solve $ \mathcal{P} 3 $. }

\subsubsection{Optimization of MUD Matrix $ \mathbf{W} $}
For given reflecting beamforming vector  $ \boldsymbol{\theta} $, $ \mathcal{P} 3 $ can be rewritten as
\begin{subequations} \label{OptF}
	\begin{align}
		\mathcal{P} 3.1: & \min _{\mathbf{W}, t} t \notag  \\
		\text { s.t. } & \dfrac{\ell_{k}}{R_{k}\left(\mathbf{w}_{k},\boldsymbol{\theta} \right) }+\dfrac{\ell_{k} c_{k}}{f_{k}^{e}} \leq t,  \forall k \in \mathcal{K}, \label{OptF1} \\
		& \left\| \mathbf{w}_{k} \right\|^{2} \leq 1, \quad  \forall k \in \mathcal{K}.  \label{OptF2}
	\end{align}
\end{subequations}
Since $ \mathcal{P} 3.1 $ is still non-convex, we introduce the following feasibility-check problem $ \mathcal{P} 3.2 $ by fixing $ t $ like $ \mathcal{P} 2.1 $.
\begin{subequations} \label{OptG}
	\begin{align}
		\mathcal{P} 3.2: ~& {\rm Find} \left\lbrace \mathbf{w}_{k}\right\rbrace \notag  \\
		\text { s.t. } & (\ref{OptF1}), (\ref{OptF2}). \label{OptG1} 
	\end{align}
\end{subequations}
Assuming that the optimal solution of problem $ \mathcal{P} 3.1 $ is $ t^{*} $, similar to the analysis in Section III A, for any given $ t \geq t^{*} $, $ t $ is a feasible solution to $ \mathcal{P} 3.2 $. Whereas, if $ t \leq t^{*} $,  $ \mathcal{P} 3.2 $ is infeasible. Thus, with the aid of a bisection search over $ t \geq 0 $, $ \mathcal{P} 3.1 $ can be solved equivalently by checking the feasibility of $ \mathcal{P} 3.2 $. 

Next, we investigate how to solve $ \mathcal{P} 3.2 $. The inequality in (\ref{OptF1}) can be rewritten as 
\begin{eqnarray} \label{eq1}
	\left(1+\frac{1}{2^{\frac{l_{k}}{B\left( t-t_{c,k}\right) }}-1}\right) p_{t}\left|\mathbf{w}_{k}^{H}\left(\mathbf{h}_{d,k}+\mathbf{G} \mathbf{\Theta} \mathbf{h}_{r,k}\right)\right|^{2} \geq \notag \\
	p_{t} \sum_{j=1}^{K}\left|\mathbf{w}_{k}^{H}\left(\mathbf{h}_{d,j}+ \mathbf{G} \mathbf{\Theta} \mathbf{h}_{r,j}\right)\right|^{2}+\sigma^{2}, \forall k \in \mathcal{K},
\end{eqnarray} 
where $ t_{c,k}={\ell_{k} c_{k}}/{f_{k}^{e}} $. According to (\ref{eq1}), it is clear that for a feasible solution  $ \left\lbrace \mathbf{w}_{k} \right\rbrace $, $ \mathcal{P} 3.2 $ is still feasible after any phase rotation. Here, we select a set of  $ \left\lbrace \mathbf{w}_{k} \right\rbrace $, satisfying
\begin{equation} \label{eq2}
	\mathbf{w}_{k}^{H}\left(\mathbf{h}_{d,k}+\mathbf{G}\mathbf{\Theta}\mathbf{h}_{r,k}\right) \geq 0, \forall k \in \mathcal{K}.
\end{equation}
Let $ \operatorname{Re}\left(x\right) $ and $\operatorname{Im}\left( x\right) $ denote the real and imaginary parts operation. The real part and imaginary part of  
 $ \mathbf{w}_{k}^{H}\left(\mathbf{h}_{d,k}+\mathbf{G}\mathbf{\Theta}\mathbf{h}_{r,k}\right) $ are  non-negative and zero, respectively, i.e., $ \operatorname{Re}\left(\mathbf{w}_{k}^{H}\left(\mathbf{h}_{d,k}+\mathbf{G}\mathbf{\Theta}\mathbf{h}_{r,k}\right)\right) \geq 0 $ and $ \operatorname{Im}\left(\mathbf{w}_{k}^{H}\left(\mathbf{h}_{d,k}+\mathbf{G}\mathbf{\Theta}\mathbf{h}_{r,k}\right)\right)=0 $.
 Then, we define a matrix $ \mathbf{A} $, in which the $ (j,k) $-th element is $ \sqrt{p_{t}}\mathbf{w}_{k}^{H}\left(\mathbf{h}_{d,j}+\mathbf{G}\mathbf{\Theta}\mathbf{h}_{r,j}\right) $. Let $ \left\|.\right\|_{2} $ denote the 2-norm of a vector, and thus (\ref{eq1}) can be reformulated as
\begin{equation} \label{eq3}
	\sqrt{1\!+\!\frac{1}{2^{\frac{l_{k}}{B\left( t-t_{c}\right) }}\!-\!1}} \mathbf{w}_{k}^{H}\left(\mathbf{h}_{d,k}\!+\!\mathbf{G}\mathbf{\Theta}\mathbf{h}_{r,k}\right) \!\geq\!\left\|\begin{array}{c}
		\!\mathbf{A}^{H}\! \mathbf{e}_{k} \!\!\\
		\sigma_{k}
	\end{array}\right\|_{2}, \\
   \forall k\! \in\! \mathcal{K}, 
\end{equation}
where $ \mathbf{e}_{k} \in \mathbb{C}^{K \times 1} $ represents a vector, in which the $ k $-th element is one and other elements are zero. Hence, for a given $ t $ at the $ l $-th iteration, $ \mathcal{P} 3.2 $ can be equivalently expressed as
\begin{subequations} \label{OptH}
	\begin{align}
	   &  \mathcal{P} 3.3:  {\rm Find: } \left\lbrace \mathbf{w}_{k}\right\rbrace \notag  \\
	   & \text { s.t. }  (\ref{OptF2}), (\ref{eq2}), \\
	   & \sqrt{1\!+\!\frac{1}{2^{\frac{l_{k}}{B\left( t^{l}-t_{c}\right) }}\!-\!1}} \mathbf{w}_{k}^{H}\!\!\left(\mathbf{h}_{d,k}\!+\!\mathbf{G}\mathbf{\Theta}\mathbf{h}_{r,k}\!\right)
	    \!\geq\!\left\|\begin{array}{c}
			\!\!\!\mathbf{A}^{H}\! \mathbf{e}_{k} \!\!\!\\
			\sigma_{k}
		\end{array}\right\|_{2}\!\!, 
	   \forall k\! \in\! \mathcal{K}.
	\end{align}
\end{subequations}
It is not difficult to observe that $ \mathcal{P} 3.3 $ is an SOCP problem and can be solved using the existing standard convex optimization techniques (i.e., CVX solver) \cite{ref37}. 

\subsubsection{Optimization of Reflecting Beamforming Vector $ \boldsymbol{\theta} $} 
Different from solving $ \mathbf{W} $, the SOCP technique cannot be used to find the optimal $ \boldsymbol{\theta} $ since $ \mathbf{\Theta} $ is common, resulting in $  \mathbf{w}_{k}^{H}\!\left(\mathbf{h}_{d,k}\!+\!\mathbf{G}\mathbf{\Theta}\mathbf{h}_{r,k}\right) $ cannot be ensured to be real numbers for all $ k \in \mathcal{K}  $. 

To proceed, we first define $ \mathbf{v}=\left[\beta_{1} e^{j \theta_{1}}, \ldots, \beta_{RN} e^{j \theta_{IN}}\right]^{T} $, $ \mathbf{a}_{k,j}=\sqrt{P_{t}}\operatorname{diag}(\mathbf{h}_{r,k}^{H})\mathbf{G}^{H}\mathbf{w}_{k} $ and $ d_{k,j}=\sqrt{P_{t}}\mathbf{w}_{k}^{H}\mathbf{h}_{d,j} $, $ \forall k, j \in \mathcal{K} $.  
Consequently, we have
\begin{equation}
P_{t}\!\left| \mathbf{w}_{k}^{H}\!\left(\mathbf{h}_{d,j}\!+\!\mathbf{G}\mathbf{\Theta}\mathbf{h}_{r,j}\!\right) \right|^{2}\!\!\!=\!\mathbf{v}^{H}\mathbf{C}_{k,j}\mathbf{v}\!+\!2\! \operatorname{Re}\!\left\{\mathbf{v}^{H} \mathbf{u}_{k, j}\right\}\!+\!\left|d_{k, j}\right|^{2}\!\!\!,
\end{equation}
where $ \mathbf{C}_{k,j}\!=\!\mathbf{a}_{k,j}\mathbf{a}_{k,j}^{H} $ and $ \mathbf{u}_{k, j}\!=\!\sqrt{P_{t}}\mathbf{w}_{k}^{H}\mathbf{h}_{d,j}\operatorname{diag}\left( \mathbf{h}_{r,j}^{H} \right)\mathbf{G}^{H}\mathbf{w}_{k} $. After obtaining $ \mathbf{W} $, $ \mathcal{P} 3 $ can be reformulated as 
\begin{subequations} \label{OptI}
	\begin{align}
		\mathcal{P} 3.4: & \min _{\mathbf{v}, t} t \notag  \\
		\text { s.t. } & \dfrac{\!\mathbf{v}^{H}\mathbf{C}_{k,k}\mathbf{v}\!+\!2 \operatorname{Re}\left\{\mathbf{v}^{H} \mathbf{u}_{k, k}\right\}\!+\!\left|d_{k, k}\right|^{2}}{\sum_{j=1,j\ne k}^{K}\mathbf{v}^{H}\mathbf{C}_{k,j}\mathbf{v}\!+\!2 \operatorname{Re}\left\{\mathbf{v}^{H} \mathbf{u}_{k, j}\right\}\!+\!\left|d_{k, j}\right|^{2}+\sigma^{2} } \notag \\ & \quad \quad\quad\quad\quad\geq  2^{\frac{\ell_{k}}{B\left(t-\frac{\ell_{k} c_{k}}{f_{k}^{e}}\right)}}-1, \forall k \in \mathcal{K}, \label{OptI1} \\
		& \left|v_{n} \right|^{2}=1 , \quad n=1,2, \ldots, IN. \label{OptI2} 
	\end{align}
\end{subequations}
$ \mathcal{P} 3.4 $ is still difficult to be solved directly due to non-convex constraints (\ref{OptI1}) and (\ref{OptI2}). Next, we develop two effective schemes based on SDR and SCA techniques to solve it.

\subsection*{Solving $ \mathcal{P} 3.4 $ Based on SDR}
To solve $ \mathcal{P} $3.4, we first reformulate it as 
\begin{subequations} \label{OptJ}
	\begin{align}
	&	\mathcal{P} 3.5:  \min _{\overline{\mathbf{v}}, t} t \notag\\
		\text { s.t. } & \frac{\overline{\mathbf{v}}^{H} \mathbf{R}_{k, k} \overline{\mathbf{v}}+\left|d_{k, k}\right|^{2}}{\sum_{j=1, j \neq k}^{K} \overline{\mathbf{v}}^{H} \mathbf{R}_{k, j} \overline{\mathbf{v}}+\left|d_{k, j}\right|^{2}+\sigma^{2}} \geq \alpha_{k}\left(t \right), \forall k \in \mathcal{K}, \\ \label{OptJ1}
		&\left|\bar{v}_{n}\right|=1, \quad n=1,2, \ldots, IN+1,  
	\end{align}
\end{subequations}
where $ \alpha_{k}\left(t \right)=2^{\frac{\ell_{k}}{B\left(t-\frac{\ell_{k} c_{k}}{f_{k}^{e}}\right)}}-1  $, $ \mathbf{R}_{k,j} $ and $ \overline{\mathbf{v}} $ are defined as
\begin{equation}
	\mathbf{R}_{k, j}=\left[\begin{array}{cc}
		\mathbf{C}_{k, j} & \mathbf{u}_{k, j} \\
		\mathbf{u}_{k, j}^{H} & 0
	\end{array}\right] \text { and } \overline{\mathbf{v}}=\left[\begin{array}{l}
		\mathbf{v} \\
		1
	\end{array}\right],
\end{equation} 
respectively. Define $\mathbf{V}=\overline{\mathbf{v}} \overline{\mathbf{v}}^{H}$, which needs to satisfy $\mathbf{V} \succeq 0$ and $\operatorname{rank}(\mathbf{V})=1$. Accordingly, $  \mathcal{P} $3.5 can be re-expressed as
\begin{subequations} \label{OptK}
	\begin{align}
		&\mathcal{P}3.6: \min _{\mathbf{V}, t} t \notag\\
		&\text { s.t. } \frac{\operatorname{Tr}\left(\mathbf{R}_{k, k} \mathbf{V}\right)+\left|d_{k, k}\right|^{2}}{\sum_{j=1, j \neq k}^{K} \operatorname{Tr}\left(\mathbf{R}_{k, j} \mathbf{V}\right)+\left|d_{k, j}\right|^{2}+\sigma_{k}^{2}} \!\geq\! \alpha_{k}\left(  t\right) , \forall k \in \mathcal{K}, \label{OptK1}\\
		&\quad \mathbf{V}_{n, n} = 1,  \quad  n=1,...,IN+1, \label{OptK2}\\
		&\quad \mathbf{V} \succeq 0, \label{OptK3}\\
		&\quad \operatorname{rank}(\mathbf{V}) = 1 \label{OptK4},
	\end{align}
\end{subequations}
where $\operatorname{Tr}(\boldsymbol{.})$ represents the trace operation.  $ \mathcal{P} 3.6 $ is non-convex due to the rank-one constraint. To proceed it, we relax this rank-one constraint and solve $ \mathcal{P} 3.6 $ by solving the following feasibility problem based on the above analysis, with a bisection search over $ t $, which is formulated as
\begin{subequations} \label{OptLL}
	\begin{align}
		\mathcal{P}3.7: &\text {Find: } \mathbf{V} \nonumber \\
	\text { s.t. } &\operatorname{Tr}\left(\mathbf{R}_{k, k} \mathbf{V}\right)+\left|d_{k, k}\right|^{2} \geq \nonumber \\
	&\alpha_{k}\left( t^{l_{2}} \right) \left(\sum_{j \neq k, j \in \mathcal{K}}\!\! \operatorname{Tr}\left(\mathbf{R}_{k, j} \mathbf{V}\right)\!+\!\left|d_{k, j}\right|^{2}\!+\!\sigma_{k}^{2}\right), \forall k \in \mathcal{K} \label{OptLL1} \\
    &\quad \mathbf{V}_{n, n} = 1,  \quad  n=1,...,IN+1 \label{OptLL2}\\
    &\quad \mathbf{V} \succeq 0, \label{OptLL3}
	\end{align}
\end{subequations}
where $ t^{l_{2} } $ is the value of $ t $ at the $ l_{2} $-th iteration. It is obvious that $ \mathcal{P}3.7 $ is a classical SDR problem, and the optimal solution can be found by CVX solver. However, the SDR may not be tight for $ \mathcal{P}3.6 $. In this case, we can use the Gaussian randomization technique \cite{ref39} to obtain a feasible solution to $ \mathcal{P}3.6 $ based on the high-rank solution obtained in $ \mathcal{P}3.7 $. 

In summary, we solve $ \mathcal{P} 3 $ by alternately solving $ \mathcal{P} 3.1 $ and $ \mathcal{P} 3.4 $, and the detailed procedure is presented in \textbf{Algorithm 2}. For each iteration, $ \mathcal{P} 3.1 $ is first solved based on obtained $ \boldsymbol{\theta} $ at the previous iteration, and then $ \mathcal{P} 3.4 $ is solved based on obtained $ \mathbf{W} $ in step 3. We start from solving $ \mathcal{P} 3.1 $ instead of  solving $ \mathcal{P} 3.4 $ because $ \mathcal{P} 3.1 $ is always feasible under any given $ \boldsymbol{\theta} $, but the reverse may not be true. Furthermore, the rigorous proof of convergence is provided in Proposition~2.

\emph{Proposition 2: The OF of $ \mathcal{P} 3 $ is monotonically non-increasing based on the proposed \textbf{Algorithm 2}.}

\emph{Proof:} See Appendix B.

\emph{Remark 3: When the solution of $ \mathcal{P} 3.7 $ is not rank-one, we need to reconstruct a rank-one solution by leveraging the Gaussian randomization procedure. Therefore, the performance of the reconstructed solution critically depends on the generated Gaussian randomization number. For example, due to the uncertainty in randomizations, we may find a highly suboptimal solution to $ \mathcal{P} 3.4 $, which will lead to a compromised performance. In addition, to find a better solution, we need to generate a large number of Gaussian randomizations. Besides, solving the SDR problem is time-consuming, especially when the matrix dimension is large. Therefore, in order to reduce the computational complexity and guarantee performance, we need to further develop an efficient algorithm. }

\begin{algorithm}[t]
	\caption{Joint optimization of $ t $, $ \mathbf{W} $, and $ \boldsymbol{\theta} $ with SDR technique}
	\begin{algorithmic}[1] 
		\Require 
		$ l_{2}=0 $, $ \boldsymbol{\theta}^{(0)} $ and accuracy threshold $ \epsilon_{2}>0 $.
		\Repeat:
		\State  Under given $ \left( t^{\left( l_{2}\right) }, \boldsymbol{\theta}^{(l_{2})} \right) $, solving $ \mathcal{P} 3.1 $ to obtain $ t^{\star} $ and $\mathbf{W}^{\star}$, and set $ t^{* \left( l_{2}\right)}=t^{\star} $, $\mathbf{W}^{(l_{2})}=\mathbf{W}^{\star} $
		\State Under given $\left( t^{* \left( l_{2}\right)},  \mathbf{W}^{(l_{2})} \right) $, solving $ \mathcal{P} 3.4 $ to obtain $ t^{\star} $ and $ \boldsymbol{\theta}^{\star} $, and set  $ t^{\left( l_{2}+1\right) }=t^{\star} $, $ \boldsymbol{\theta}^{\left( l_{2}+1\right) }=\boldsymbol{\theta}^{\star} $
		\State Update $ l_{2} \gets l_{2}+1 $
		\Until{ Convergence }
	\end{algorithmic}
\end{algorithm}

\subsection*{Solving $ \mathcal{P} 3.4 $ Based on SCA}
To overcome the drawbacks of the SDR technique, we propose an efficient scheme to update the reflecting beamforming vector $ \mathbf{v} $ based on the SCA technique. Instead of solving $ \mathcal{P} 3.4 $ directly, we try to find a feasible solution $ \mathbf{v} $ to reduce the maximum edge computing latency. Particularly, for a specific iteration $ l \geq 1 $, we first define $ \mathbf{v}^{\left( l-1\right) } $ as the value of $ \mathbf{v} $ obtained at the previous iteration. Then, for given $ \left(\left\lbrace \mathbf{w}_{k}^{l}\right\rbrace, \mathbf{v}^{\left( l-1\right) } \right)  $, the achieved maximum WD's edge computing latency is denoted by $ t^{(l)}={\max}_{k \in \mathcal{K}}  D_{k}^{e}\left(\left\{\mathbf{w}_{k}^{l}\right\}, \mathbf{v}^{\left( l-1\right) }\right) $. 

Firstly, according to ($ \ref{OptI1} $), we introduce an auxiliary function $ \mathcal{F}_{k} $, which is defined as 
\begin{small}
\begin{equation}\label{eq5}
	\begin{aligned}
		&\mathcal{F}_{k}\left(\mathbf{v},\left\{\mathbf{w}_{k}\right\}, t\right) \\
		&=\alpha_{k}\left(  t\right)\! \left[\sum_{j \neq k, j \in \mathcal{K}}\!\!\left(\mathbf{v}^{H} \mathbf{C}_{k, j} \mathbf{v}\!+\!2 \operatorname{Re}\left\{\mathbf{v}^{H} \mathbf{u}_{k, j}\right\}\!+\!\left|d_{k, j}\right|^{2}\right)\!+\!\sigma_{k}^{2}\right] \\
		&\quad-\left(\mathbf{v}^{H} \mathbf{C}_{k, k} \mathbf{v}+2 \operatorname{Re}\left\{\mathbf{v}^{H} \mathbf{u}_{k, k}\right\}+\left|d_{k, k}\right|^{2}\right), \forall k \in \mathcal{K}.
	\end{aligned}
\end{equation}
\end{small}
\!\!\!If $ \left( \mathbf{v},\left\{\mathbf{w}_{k}\right\}, t \right)  $ is a set of feasible solutions to $ \mathcal{P} 3.4 $, we have $ \mathcal{F}_{k} \leq 0 $, $\forall k \in \mathcal{K} $. After solving $ \mathcal{P} 3.1 $ at each iteration, the maximum $ \mathcal{F}_{k} $ at WDs is equal to 0, i.e., $ \max _{k \in \mathcal{K}} \mathcal{F}_{k}\left(\mathbf{v}^{(l-1)},\left\{\mathbf{w}_{k}^{(l)}\right\}, t^{(l)}\right)=0 $. Consequently, the reflecting beamforming vector can be updated by equivalently solving the following optimization problem
\begin{equation}
	\begin{gathered}
		\mathcal{P}3.8: \min _{\mathbf{v}} \max _{k \in \mathcal{K}} \mathcal{F}_{k}\left(\mathbf{v},\left\{\mathbf{w}_{k}^{(l)}\right\}, t^{(l)}\right) \\
		\text { s.t. } (\ref{OptI2}).
	\end{gathered}
\end{equation}
When $\max _{k \in \mathcal{K}} \mathcal{F}\left(\mathbf{v}^{(l)},\left\{\mathbf{w}_{k}^{(l)}\right\}, t^{(l)}\right) < 0$, it can be easily proved that $\max _{k \in \mathcal{K}} D_{k}^{e}\left(\mathbf{v}^{(l)},\left\{\mathbf{w}_{k}^{(l)}\right\}\right)<\max _{k \in \mathcal{K}} D_{k}^{e}\left(\mathbf{v}^{(l-1)},\left\{\mathbf{w}_{k}^{(l)}\right\}\right)$, i.e., the maximum edge computing latency is reduced. Thus, $ \mathcal{P} 3.4 $ can be equivalently solved by solving $ \mathcal{P} 3.8 $. Next, we investigate how to solve $ \mathcal{P} 3.8 $.

Since the OF of $ \mathcal{P} 3.8 $ is non-convex, it is difficult to solve $ \mathrm{P} 3.8 $ directly. Inspired by the SCA technique, we first obtain a convex upper bound on $ \mathcal{F}_{k} $ by approximating the second convex term based on its first-order Taylor expansion. Therefore, under given $ t^{\left( l \right) } $, $ \left\{ \mathbf{w}_{k}^{(l)} \right\} $, and local point $ \mathbf{v}^{(l-1)} $, the lower bounded of  $ \mathcal{F}_{k} $ is
\begin{small}
\begin{equation}
	\begin{aligned}
		&\mathcal{F}_{k}\left(\mathbf{v},\left\{\mathbf{w}_{k}^{(l)}\right\}, t^{(l)}\right) \leq \\
		&\quad \alpha_{k}\!\left(  t^{(l)}\right) \!\left[\sum_{j \neq k, j \in \mathcal{K}}\!\!\left(\mathbf{v}^{H} \mathbf{C}_{k,j} \mathbf{v}\!+\!2 \operatorname{Re}\left\{\mathbf{v}^{H} \mathbf{u}_{k, j}\right\}\!+\!\left|d_{k, j}\right|^{2}\right)\!+\!\sigma_{k}^{2}\right] \\
		&\quad-\left(\mathbf{v}^{(l-1)^{H}} \mathbf{C}_{k, k} \mathbf{v}^{(l-1)}+2 \operatorname{Re}\left\{\mathbf{v}^{(l-1)^{H}} \mathbf{u}_{k, k}\right\}+\left|d_{k, k}\right|^{2}\right) \\
		&\quad-2\left(\mathbf{C}_{k, k}^{H} \mathbf{v}^{(l-1)}+\mathbf{u}_{k, k}\right)^{H}\left(\mathbf{v}-\mathbf{v}^{(l-1)}\right) \\
		&\triangleq \mathcal{F}_{k}^{\mathrm{up}}\left(\mathbf{v},\left\{\mathbf{w}_{k}^{(l)}\right\}, t^{(l)}, \mathbf{v}^{(l-1)}\right).
	\end{aligned}
\end{equation}
\end{small}
Next, we replace $ \mathcal{F}_{k}\left(\mathbf{v},\left\{\mathbf{w}_{k}^{(l)}\right\}, t^{(l)}\right) $ by $ \mathcal{F}_{k}^{\mathrm{up}}\left(\mathbf{v},\left\{\mathbf{w}_{k}^{(l)}\right\}, t^{(l)}, \mathbf{v}^{(l-1)}\right) $ and introduce another auxiliary variable $ z $, $ \mathcal{P} 3.8 $ can be approximated as
\begin{equation}
	\begin{aligned}
     (\mathcal{P} 3.9): & \min _{\mathbf{v}, z} z\\
     \text { s.t. } & \mathcal{F}_{k}^{\mathrm{up}}\left(\mathbf{v},\left\{\mathbf{w}_{k}^{(l)}\right\}, t^{(l)}, \mathbf{v}^{(l-1)}\right) \leq z, \forall k \in \mathcal{K}, \\
     & (\ref{OptI2}).
	\end{aligned}
\end{equation}
So far, we have transformed $ \mathcal{P} 3.8 $ into a convex form, which can be solved by existing convex optimization solvers such as CVX \cite{ref37}. Suppose that $\mathbf{v}^{*}$ is the optimal solution to $ \mathcal{P} 3.9 $. By substituting ($ \mathbf{v}^{*}, \left\{\mathbf{w}_{k}^{(l)}\right\}, t^{(l)} $) into $ \mathcal{P} 3.8 $, it is observed that the feasible solution of $ \mathcal{P} 3.9 $ is still feasible to $ \mathcal{P} 3.8 $, and thus solving $ \mathcal{P} 3.9 $ can achieve a smaller value than solving $ \mathcal{P} 3.8 $. The details of the SCA-based algorithm are presented in \textbf{Algorithm 3}. 

\begin{algorithm}[t]
	\caption{ Joint optimization of $ t $, $ \mathbf{W} $ and $ \boldsymbol{\theta} $ with SCA  technique}
	\begin{algorithmic}[1] 
        \Require 
		 Initialize the phase shift vector $ \mathbf{v}^{(0)} $, set 
		$ l_{3}=1 $ and accuracy threshold $ \epsilon_{3}>0 $.
		\Repeat:
		\State  Under given $ \left( t^{(l_{3}-1)}, \mathbf{v}^{(l_{3}-1)} \right) $, solve $ \mathcal{P} 3.1 $ to obtain $ t^{\star} $ and $ \mathbf{W}^{\star} $, and set $ t^{\left( l_{3}\right) }=t^{\star} $,  $ \mathbf{W}^{\left( l_{3}\right) }=\mathbf{W}^{\star} $
		\State Under given $ \left( t^{\left( l_{3}\right) }, \mathbf{W}^{\left( l_{3}\right) } \right) $, solve $ \mathcal{P} 3.9 $ to update $ \mathbf{v} $, and set $ \mathbf{v}^{l_{3}}=\mathbf{v}^{\star} $ 
	    \State Update $ l_{3} \gets l_{3}+1 $
		\Until{ Convergence }
	\end{algorithmic}
\end{algorithm}

\subsection{Overall Algorithm to Solve $ \mathcal{P} 1 $}
In \textbf{Algorithm 4}, we provide the details of the BCD-based algorithm for solving $ \mathcal{P} 1 $. It is worth noting that the auxiliary variable $ t $, which is also the OF of $ \mathcal{P} 1 $, is constantly updated with the implementation of steps 2 and 3. Let $ g\left( \boldsymbol{\ell}^{(l_{4}-1)}, \mathbf{f}^{e (l_{4}-1)}, \mathbf{W}^{(l_{4}-1) }, \boldsymbol{\theta}^{(l_{4}-1) }\right)  $ and $ g\left( \boldsymbol{\ell}^{\left( l_{4}\right) }, \mathbf{f}^{e \left( l_{4}\right) }, \mathbf{W}^{(l_{4}-1)}, \boldsymbol{\theta}^{(l_{4}-1)}\right)  $ denote the value of $ t $ after implementing step 3 at the $ (l_{4}-1) $-th iteration, and the value of $ t $ after implementing step 2 at the $ l_{4} $-th iteration, respectively. It follows that $ g\left( \boldsymbol{\ell}^{\left( l_{4}\right) }, \mathbf{f}^{e \left( l_{4}\right) }, \mathbf{W}^{(l_{4}-1)}, \boldsymbol{\theta}^{(l_{4}-1)}\right) \geq g\left( \boldsymbol{\ell}^{(l_{4}-1)}, \mathbf{f}^{e (l_{4}-1)}, \mathbf{W}^{(l_{4}-1)}, \boldsymbol{\theta}^{(l_{4}-1) }\right) $, this is due to that in step 2, $t=g\left( \boldsymbol{\ell}^{\left( l_{4}\right) }, \mathbf{f}^{e \left( l_{4}\right) }, \mathbf{W}^{(l_{4}-1)}, \boldsymbol{\theta}^{(l_{4}-1)}\right)  $ can be regarded as the overall maximum WD's latency, while in step 3, $ t=g\left( \boldsymbol{\ell}^{\left( l_{4}\right) }, \mathbf{f}^{e \left( l_{4}\right) }, \mathbf{W}^{\left( l_{4}\right)   }, \boldsymbol{\theta}^{\left( l_{4} \right)  }\right)  $ refers to the maximum WD's latency for edge computing. However, the convergence of \textbf{Algorithm 4} can be guaranteed due to the WD's overall latency and WD's edge processing latency are both reduced at different iterations. Hence, we have $ g\left( \boldsymbol{\ell}^{\left( l_{4}\right) }, \mathbf{f}^{e \left( l_{4}\right) }, \mathbf{W}^{\left(l_{4}-1 \right) }, \boldsymbol{\theta}^{\left(l_{4}-1 \right)}\right) \geq g\left( \boldsymbol{\ell}^{(l_{4}-1)}, \mathbf{f}^{e (l_{4}-1)}, \mathbf{W}^{(l_{4}-1)}, \boldsymbol{\theta}^{(l_{4}-1)}\right) \geq  g\left( \boldsymbol{\ell}^{\left( l_{4}\right) }, \mathbf{f}^{e \left( l_{4}\right) }, \mathbf{W}^{ \left( l_{4} \right)  }, \boldsymbol{\theta}^{\left( l_{4} \right)  }\right)$.
\renewcommand{\algorithmicrequire}{\textbf{Input:}}
\begin{algorithm*}[t]
	\caption{ Joint Optimization of $ \boldsymbol{\ell} $, $ \mathbf{f}^{e} $, $ \mathbf{W} $ and $ \boldsymbol{\theta $}}
	\begin{algorithmic}[0] 
		\Require 
		$ \mathbf{h}_{k} $,  $ B $, $ p_{t} $, $ \sigma^{2} $, $ L_{k} $, $ c_{k} $, $ K $, $ f_{total}^{e} $, and $ \epsilon_{4} $.
		\Ensure
		Optimal $ \boldsymbol{\ell} $, $ \mathbf{f}^{e} $, $ \mathbf{W} $ and $ \boldsymbol{\theta $}
		\State \textbf{1. Initialization}\\
		Initialize $ l_{4}=1 $, $ \epsilon_{4}^{\left( 0 \right) } \geq 0 $\\
		Initialize  $ \mathbf{W}^{\left(0 \right) } $ satisfying (\ref{OptA5}), $ \boldsymbol{\theta }^{\left( 0\right) } $ satisfying (\ref{OptA1}), and calculate $  R_{k} $ according to ($ \ref{Eq01} $), $ \forall k \in \mathcal{K}$
		\State \textbf{2. Joint optimization of $ t $, $ \boldsymbol{\ell} $ and $ \mathbf{f}^{e} $, given  $ \mathbf{R}^{\left(l_{4}-1 \right) }  $ }
		\State Calculate $\mathbf{f}^{e\left(l_{4}\right)}$ and $ \boldsymbol{\ell}^{\left( l_{4}\right) } $ and update $ t $ relying on \textbf{Algorithm 1}
		\State \textbf{3. Joint optimization of $ t $, $ \mathbf{W} $ and $ \boldsymbol{\theta } $, given  $ \boldsymbol{\ell}^{\left(l_{4} \right) } $ and $\mathbf{f}^{e\left(l_{4}\right)}$}
		\State Calculate $ \mathbf{W}^{\left(l_{4}\right) } $ and $ \boldsymbol{\theta }^{\left(l_{4} \right) } $ and update $ t $ relying on the \textbf{Algorithm 2} or \textbf{Algorithm 3}
		\State \textbf{4. Convergence checking}\\
		$\epsilon_{4}^{\left(l_{4}\right)}=\frac{\left|\operatorname{obj}\left(\boldsymbol{\ell}^{\left(l_{4}\right)}, \mathbf{f}^{e\left(l_{4}\right)}, \mathbf{w}^{\left(l_{4}\right)}, \boldsymbol{\theta}^{\left(l_{4}\right)}\right)-\operatorname{obj}\left(\boldsymbol{\ell}^{\left(l_{4}-1\right)}, \mathbf{f}^{e\left(l_{4}-1\right)}, \mathbf{w}^{\left(l_{4}-1\right)}, \boldsymbol{\theta}^{\left(l_{4}-1\right)}\right)\right|}{\operatorname{obj}\left(\boldsymbol{\ell}^{\left(l_{4}\right)}, \mathbf{f}^{e\left(l_{4}\right)}, \mathbf{w}^{\left(l_{4}\right)}, \boldsymbol{\theta}^{\left(l_{4}\right)}\right)}$
		\If{$\epsilon_{4}^{\left(l_{4}\right)}>\epsilon_{4} ~\& \& ~ l_{4}<l_{4}^{\max }$ hold} 
		   \State Calculate $ \mathbf{R}^{\left( l_{4} \right) } $ using (\ref{Eq01})
		   \State $ l_{4} \gets l_{4}+1 $
		   \State Go to Step 2
		\Else 
		  \State Output $ \boldsymbol{\ell} $, $ \mathbf{f}^{e} $, $ \mathbf{W} $ and $ \boldsymbol{\theta } $
		\EndIf
	\end{algorithmic}
\end{algorithm*}

\section{The Single WD Scenario}
To further investigate the beneficial role of IRS in the cell-free MEC system, we consider a single WD scenario in this section. Compared with the multi-WD scenario, the single-user scenario is more tractable. This is because a) the min-max latency optimization problem $ \mathcal{P}0 $ becomes a simple latency minimization problem, which is more tractable, b) there is no involved computing resource allocation, c) there is no multi-WD interference. Thus, the overall latency is reformulated as
\begin{equation}
	D( \mathbf{w}, \boldsymbol{\theta}, \ell)=\max \left\{\frac{(L-\ell) c}{f^{l}}, \frac{\ell}{R(\mathbf{w}, \boldsymbol{\theta})}+\frac{\ell c}{f_{\text {total }}^{e}}\right\}.
\end{equation}
On this basis, $ \mathcal{P}0 $ is simplified to
\begin{subequations} \label{OptM}
	\begin{align}
		\mathcal{P}4:   & \min _{\ell, \mathbf{w}, \boldsymbol{\theta}}   D(\mathbf{w}, \boldsymbol{\theta}, \ell) \notag \\
		\text { s.t. } & 0 \leq \theta_{n}<2 \pi, \quad n=1,2, \ldots, IN, \label{OptM1}\\
		& \ell \in\{0,1, \ldots, L\}, \label{OptM2}\\
		&\left\| \mathbf{w} \right\|^{2} \leq 1. \label{OptM3}
	\end{align}
\end{subequations}
Like Proposition 1, fixing $ \mathbf{w} $ and $ \boldsymbol{\theta} $, $ D(\mathbf{w}, \boldsymbol{\theta}, \ell) $ reaches its minimum threshold when $ \ell $ is chosen to ensure that local computing latency is equivalent to edge compting latency, i.e., $\frac{(L-\ell) c}{f^{l}}=\frac{\ell}{R(\mathbf{w}, \boldsymbol{\theta})}+\frac{\ell c}{f_{\text {total }}^{e}}$. Thus, the optimal offloading data size is given by
\begin{equation}\label{Eq41}
	\hat{\ell}^{*}=\frac{L c R f_{\text {total }}^{e}}{f_{\text {total }}^{e} f^{l}+c R\left(f_{\text {total }}^{e}+f^{l}\right)}.
\end{equation}
Based on obtaining $ \ell $, $ \mathcal{P}4 $ can be equivalently transformed into
\begin{subequations}
	\begin{align}
	\mathcal{P}4.1: &\min _{\mathbf{w}, \boldsymbol{\theta}} \frac{\ell}{R(\mathbf{w}, \boldsymbol{\theta})}+\frac{\ell c}{f_{\text {total }}^{e}} \notag \\
	\text { s.t. } & 0 \leq \theta_{n}<2 \pi, \quad n=1,2, \ldots, IN, \\
	&\left\| \mathbf{w} \right\|^{2} \leq 1.	
    \end{align}
\end{subequations}
Based on (\ref{Eq02}) and (\ref{Eq01}), $ \mathcal{P}4.1 $ can be simplified to
\begin{subequations} \label{OptN}
	\begin{align}
		\mathcal{P}4.2: & \max _{\mathbf{w}, \boldsymbol{\theta}} \frac{p_{t}\left|\mathbf{w}^{H}\left(\mathbf{h}_{d}+\mathbf{G}\mathbf{\Theta}\mathbf{h}_{r}\right)\right|^{2}}{\sigma^{2}\left\|\mathbf{w}^{H}\right\|^{2}} \notag\\
	    \text { s.t. } & 0 \leq \theta_{n}<2 \pi, \quad n=1,2, \ldots, IN. \label{OptN1} \\
		&\left\| \mathbf{w} \right\|^{2} \leq 1.    
	\end{align}
\end{subequations}
Next, we jointly optimize $ \mathbf{w} $ and $ \boldsymbol{\theta} $ by using the BCD technique. Specifically, fixing $ \boldsymbol{\theta} $, the maximum ratio combining (MRC) technique can be applied to find optimal $ \mathbf{w} $, which is formulated as
\begin{equation}
	\mathbf{w}=\sqrt{p_{t}}\left(\mathbf{h}_{d}+\mathbf{G}\mathbf{\Theta}\mathbf{h}_{r}\right) / \sigma.
\end{equation}
Then, after obtaining $ \mathbf{w} $, we optimize the reflecting beamforming vector $ \boldsymbol{\theta} $. The SINR is upper bounded by
\begin{equation}\label{Eq37}
	\frac{p_{t}\left|\mathbf{w}^{H}\left(\mathbf{h}_{d}+\mathbf{G}\mathbf{\Theta}\mathbf{h}_{r}\right)\right|^{2}}{\sigma^{2}\left\|\mathbf{w}^{H}\right\|^{2}} \leq \frac{p_{t}\left|\mathbf{w}^{H} \mathbf{h}_{d}\right|^{2}}{\sigma^{2}\left\|\mathbf{w}^{H}\right\|^{2}}+\frac{p_{t}\left|\mathbf{w}^{H} \mathbf{G}\mathbf{\Theta}\mathbf{h}_{r}\right|^{2}}{\sigma^{2}\left\|\mathbf{w}^{H}\right\|^{2}},
\end{equation}
The equality in (\ref{Eq37}) is achieved when $ \boldsymbol{\theta} $ satisfies 
$\arg \left\{\mathbf{w}^{H} \mathbf{h}_{d}\right\}=\arg \left\{\mathbf{w}^{H} \mathbf{G}\mathbf{\Theta}\mathbf{h}_{r}\right\}$. Thus, the optimal $ \boldsymbol{\theta} $ is given by
\begin{equation}
	\boldsymbol{\theta}=\arg \left\{\mathbf{w}^{H} \mathbf{h}_{d}\right\}-\arg \left\{\operatorname{diag}\left\{\mathbf{w}^{H} \boldsymbol{G}\right\} \mathbf{h}_{r}\right\}.
\end{equation}

\section{Simulation Results} 
In this section, we evaluate the performance of the proposed algorithms by simulation. Similar to \cite{ref40}, we consider a three-dimensional system model, which is presented in Fig. 3. Five BSs simultaneously serve two WDs, and WDs can choose to offload a portion of the computing data to MEC nodes for remote execution with the help of two IRSs. The detailed location settings for BSs and IRSs are provided in the “Location model” block of Table II.
	\begin{table}
	\centering
	\caption{Default Simulation Parameter Setting}
	\begin{tabular}{|l|r|}
		\hline
		Description & Parameter and Value\cr
		\hline  
		\hline  
		Location model & \tabincell{r}{BS: ($ 40(i-1) $ m, -200m, 3m) \\ IRS: (60m, 10m, 6m), (100m, 10m, 6m)} \\ 
		\hline
		Communication model & \tabincell{r}{Bandwidth=1 MHz\\  $ \kappa_{UB}=4.6 $, $ \kappa_{UI}=2.2 $, $ \kappa_{IB}=2.8 $  \\ $ K=2 $, $ R=2 $, $ B=5 $, $ M=2 $, $ N=10 $ \\ $ p_{t}=1 $mW \\ $ \sigma^{2}=3.98 \times 10^{-12} $ mW} \\
		\hline
		Computing model & \tabincell{r}{$ L_{k}=\left[250, 350 \right] $ Kb \\ $ c_{k}=\left[700,800 \right]  $ cycle/bit \\ $ f_{k}^{l}=\left[ 4,6\right] \time 10^{8} $ cycle/s \\  $ f_{total}^{e}=50 \times 10^{9} $ cycle/s} \\
		\hline
	\end{tabular}
\end{table}

The distance-dependent path loss model is given by \cite{ref41}
\begin{equation}\label{Eq55}
	L\left( d \right)=C_{0}\left( \frac{d}{d_{0}} \right) ^{-\kappa},  	
\end{equation}	
where $ C_{0}=-30$dB refers to the path loss at a reference distance $ d_{0}=1 $m, $ d $ stands for the individual channel distance, $ \kappa $ denotes the path loss exponent. The specific setting is presented in the “Communication model” block of Table II. Then, for the small-scale fading, we consider a Rician fading channel model for all involved channels. Thus, the channel model $ \mathbf{H} $ is given by
\begin{equation}
	\mathbf{H}=\sqrt{\frac{\beta_{\mathrm{UB}}}{1+\beta_{\mathrm{UB}}}} \mathbf{H}^{\mathrm{LoS}}+\sqrt{\frac{1}{1+\beta_{\mathrm{UB}}}} \mathbf{H}^{\mathrm{NLoS}},
\end{equation}
where $ \beta_{\mathrm{UB}} $ refers to the Rician factor, $ \mathbf{H}^{\mathrm{LoS}} $ and $ \mathbf{H}^{\mathrm{NLoS}} $ denote the LoS deterministic components and the non-LOS Rayleigh fading components, respectively. $ \mathbf{H} $ is equivalent to a Rayleigh fading channel when $ \beta_{\mathrm{UB}}=0 $ and a LoS channel when $ \beta_{\mathrm{UB}} \rightarrow \infty $. Note that for the WD-BS channel model, we need to multiply the square root of the large-scale fading coefficient by the elements of the small scale fading coefficient. Similarly, the WD-IRS and IRS-BS channels can also be generated by following the above procedure with  $ \beta_{UI} $ and $ \beta_{IB} $ denoting the Rician factors of them. Furthermore, we set $ \beta_{IB}\rightarrow \infty $, $ \beta_{UI}=0  $, and $ \beta_{UB}=0 $ as \cite{ref41}. The default settings of these parameters are specified in the “Communications model” block of Table II. Besides, the computing settings follow \cite{ref42}, and are specified in the “Computing model” block of Table II.
\begin{figure*}[t]
	\begin{minipage}[t]{0.42\textwidth}
		\centering
		\includegraphics[width=1.2\textwidth]{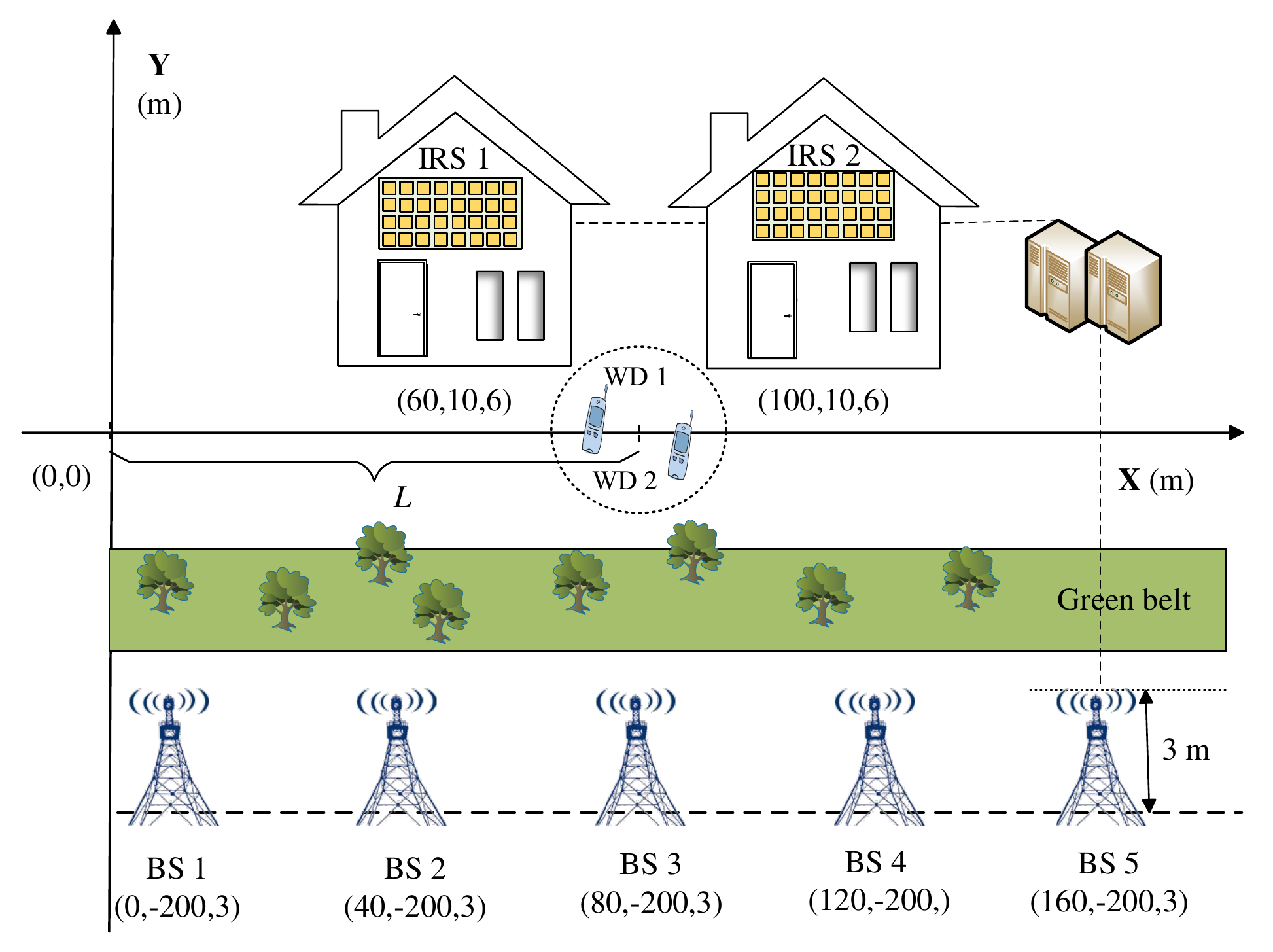}
		\caption{The simulation scenario where two WDs offload data to five BSs with the assistance of two IRSs.}
	\end{minipage}
	\qquad \qquad
	\begin{minipage}[t]{0.42\textwidth}
		\centering
		\includegraphics[width=1.2\textwidth]{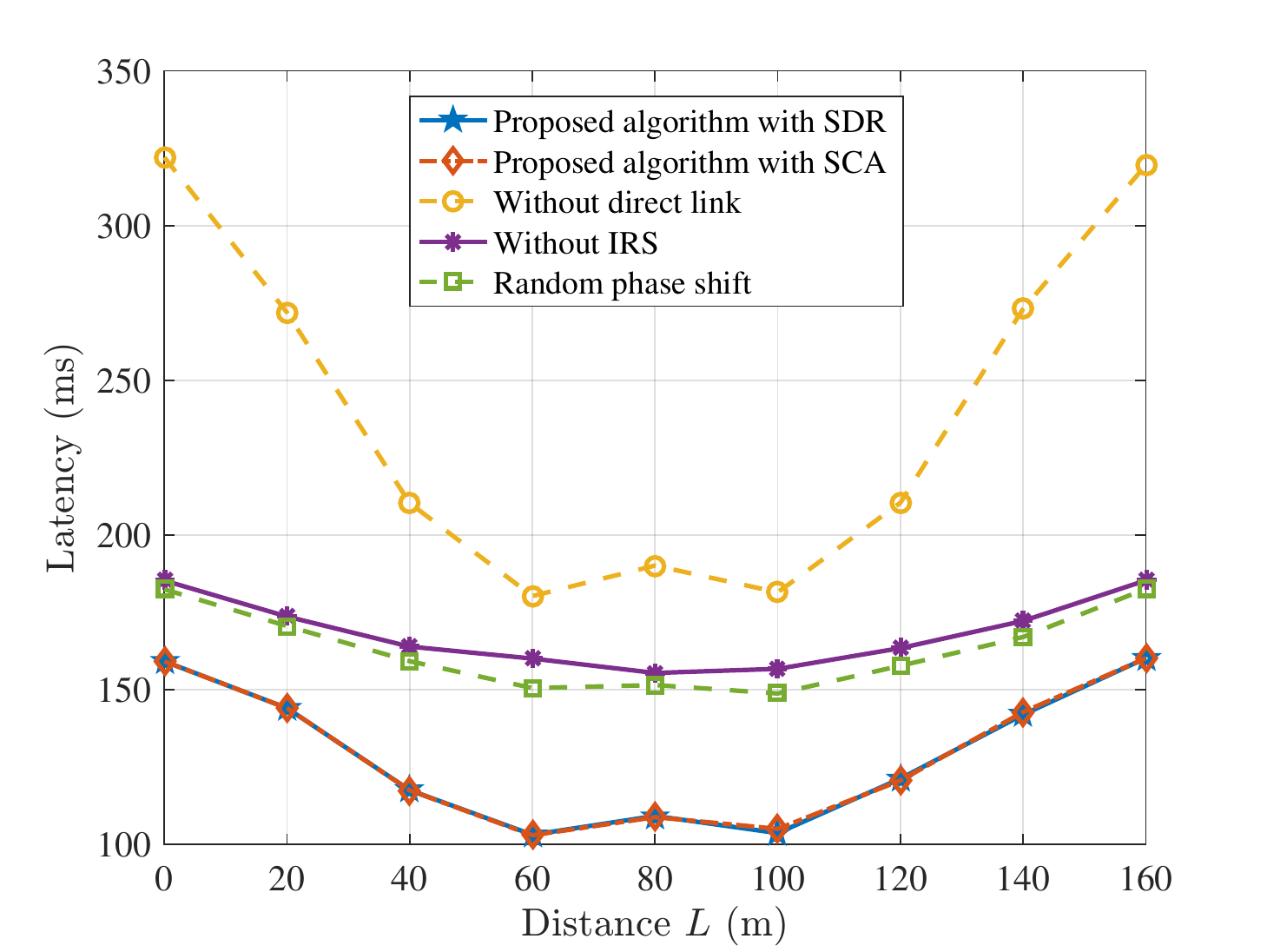}
		\caption{Maximum latency against the distance $ L $.}
	\end{minipage}
\end{figure*}
In the following, we provide simulation results to evaluate the maximum WD's latency achieved by our proposed BCD-based algorithm in various simulation environments, and compare them with the following baseline algorithms:
\begin{itemize}
	\item \emph{Without IRS: } The reflection matrix $ \mathbf{\Theta} $ is set to a zero matrix, while the other optimization variables are optimized using \textbf{Algorithm 4}.
	\item \emph{Without direct link:} Assume that all direct links between WDs and BSs are completely blocked by some moving or static object,  i.e., $ \mathbf{H} = 0 $, then, we optimize all variables using \textbf{Algorithm 4}.
	\item \emph{Random phase shift:} The IRS phase shift is set as a uniformly distributed random value in the range of $ \left[ 0, 2\pi \right)  $, while other optimization variables are optimized as \textbf{Algorithm 4}.
\end{itemize}

Figs. 4-8 show the latency under different parameter settings.

\emph{1) Impact of WDs' Locations:}  Fig. 4 shows the latency achieved under different algorithms/schemes versus WDs' locations. It is observed that, compared to baseline algorithms, our proposed two algorithms can achieve a lower latency, especially when WDs are close to IRSs. Besides, for all considered schemes with IRS, there are two obvious valleys at $ L = 60 $ m and $ L = 100 $ m. This is because IRSs can receive stronger signals transmitted from WDs when WDs approach either of two IRSs. Additionally, we observe that the latency increases when WDs are far away from BSs under the “Without IRS” scheme. Compared to “Without IRS” scheme, the reduced latency by the “Random phase shift” scheme is very limited. Furthermore, it is observed that “Without direct link” scheme achieves the highest latency, and the latency gap is larger than that of the proposed schemes when WDs approach one of two IRSs. Therefore, we conclude that deploying IRSs with optimized phase shift design can extend the signal coverage, improve signal transmission environment, and reduce latency.

\emph{2) Impact of Edge Computing Capability:} Fig. 5 shows the latency versus the edge computing capability $ f_{total}^{e} $  under different schemes. It is observed that, for all schemes, when $ f_{total}^{e} $ is small, the latency reduces drastically with the increase of $ f_{total}^{e} $, while the latency reduces slowly when $ f_{total}^{e} $ reaches a certain value, i.e., $ 30 \times 10^{9} $ cycle/s. This is because when $ f_{total}^{e} $ is small, the edge processing latency plays a dominant role, while when $ f_{total}^{e} $ becomes large, the offloading latency dominates. This indicates that, to minimize latency, adding appropriate computing capability at edge server is cost-effective.
\begin{figure*}[t]
	\begin{minipage}[t]{0.42\textwidth}
		\centering
		\includegraphics[width=1.2\textwidth]{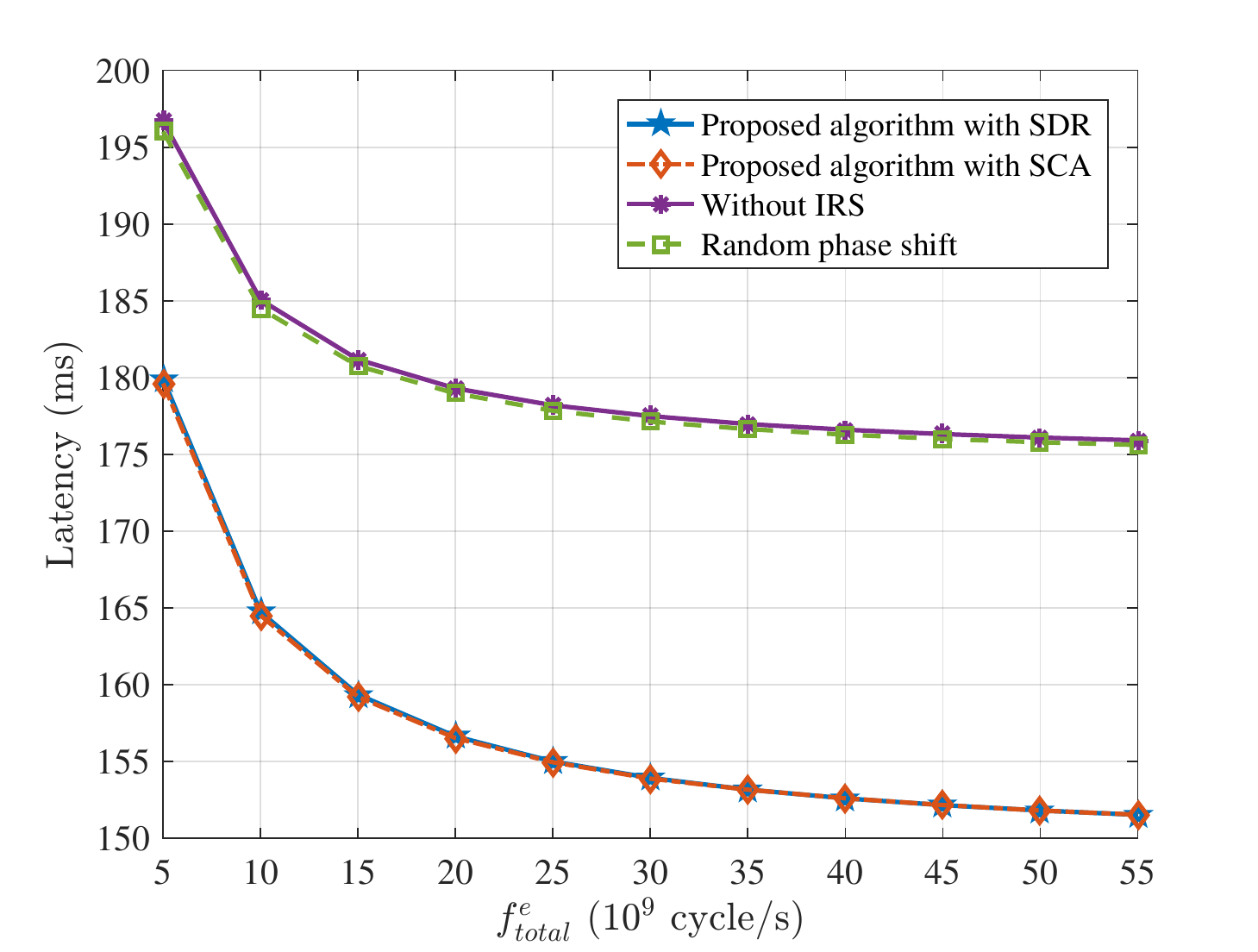}
		\caption{Maximum latency against the edge computing capability $ f^{e}_{total} $.}
	\end{minipage}
	\qquad \qquad
	\begin{minipage}[t]{0.42\textwidth}
		\centering
		\includegraphics[width=1.2\textwidth]{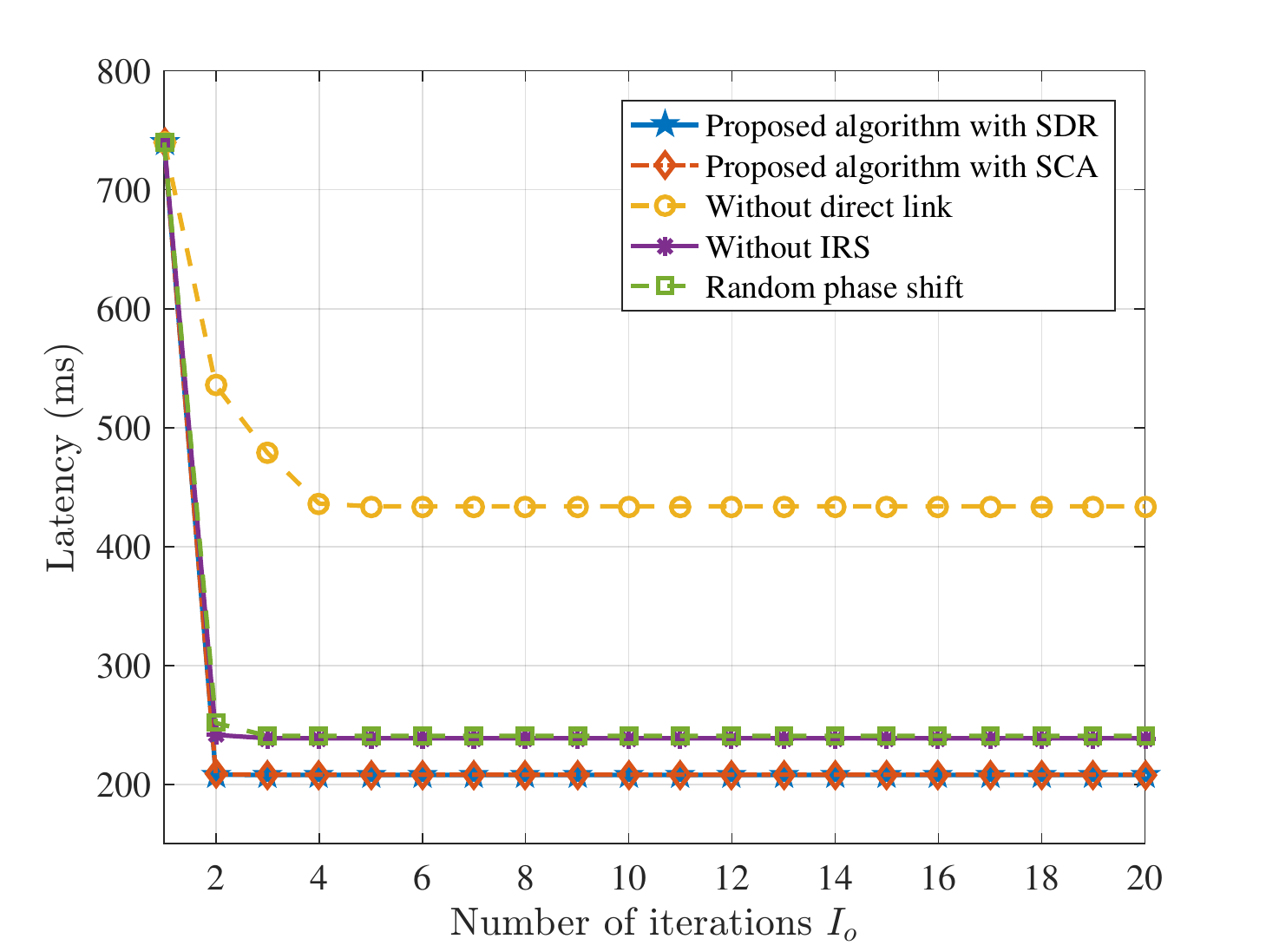}
		\caption{Maximum latency against the number of iterations $ I_{o} $.}
	\end{minipage}
\end{figure*}

\begin{figure*}[t]
	\begin{minipage}[t]{0.42\textwidth}
		\centering
		\includegraphics[width=1.2\textwidth]{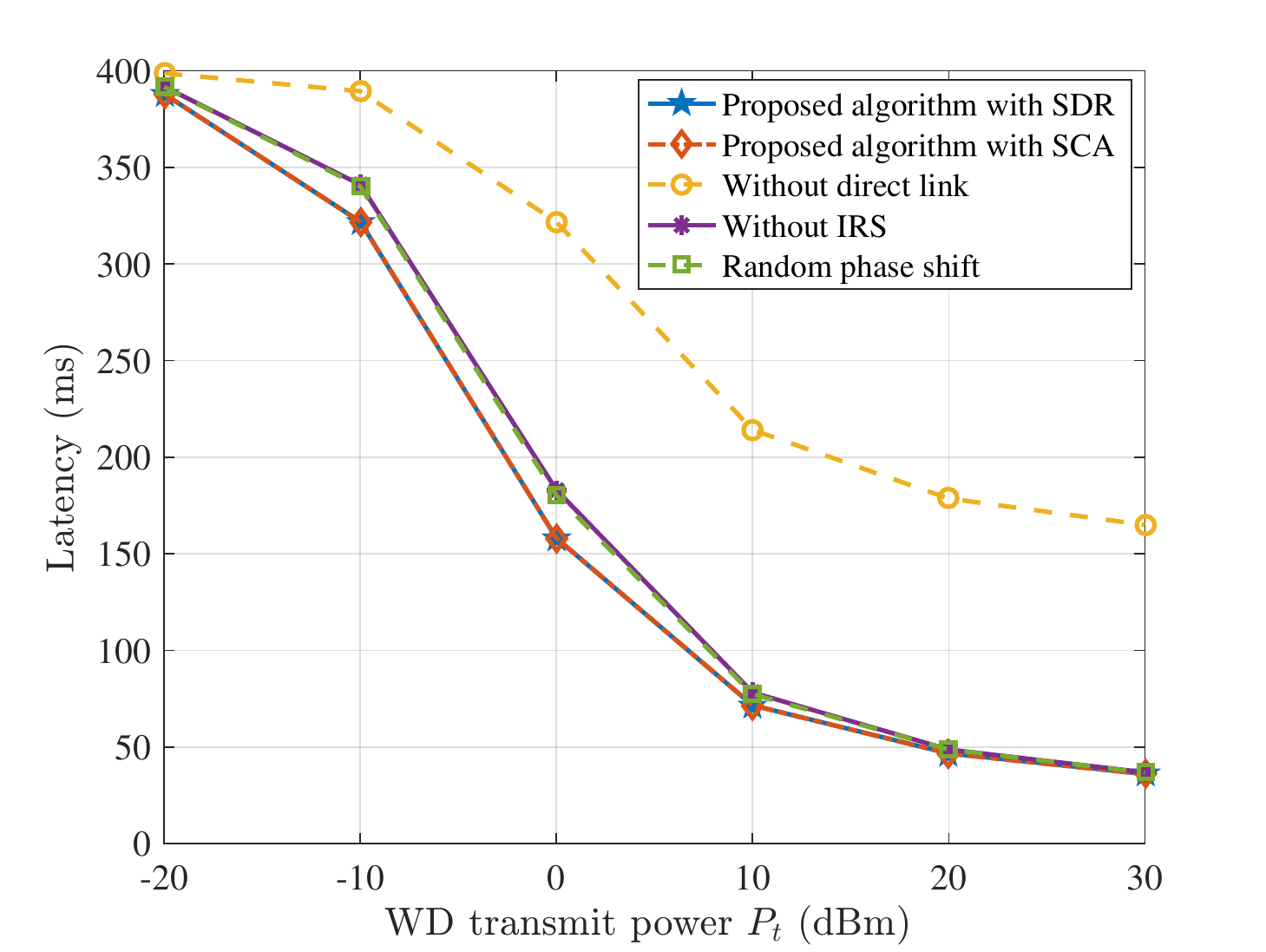}
		\caption{Maximum latency against the WDs' transmit power $ p_{t} $.}
	\end{minipage}
	\qquad \qquad
	\begin{minipage}[t]{0.42\textwidth}
		\centering
		\includegraphics[width=1.2\textwidth]{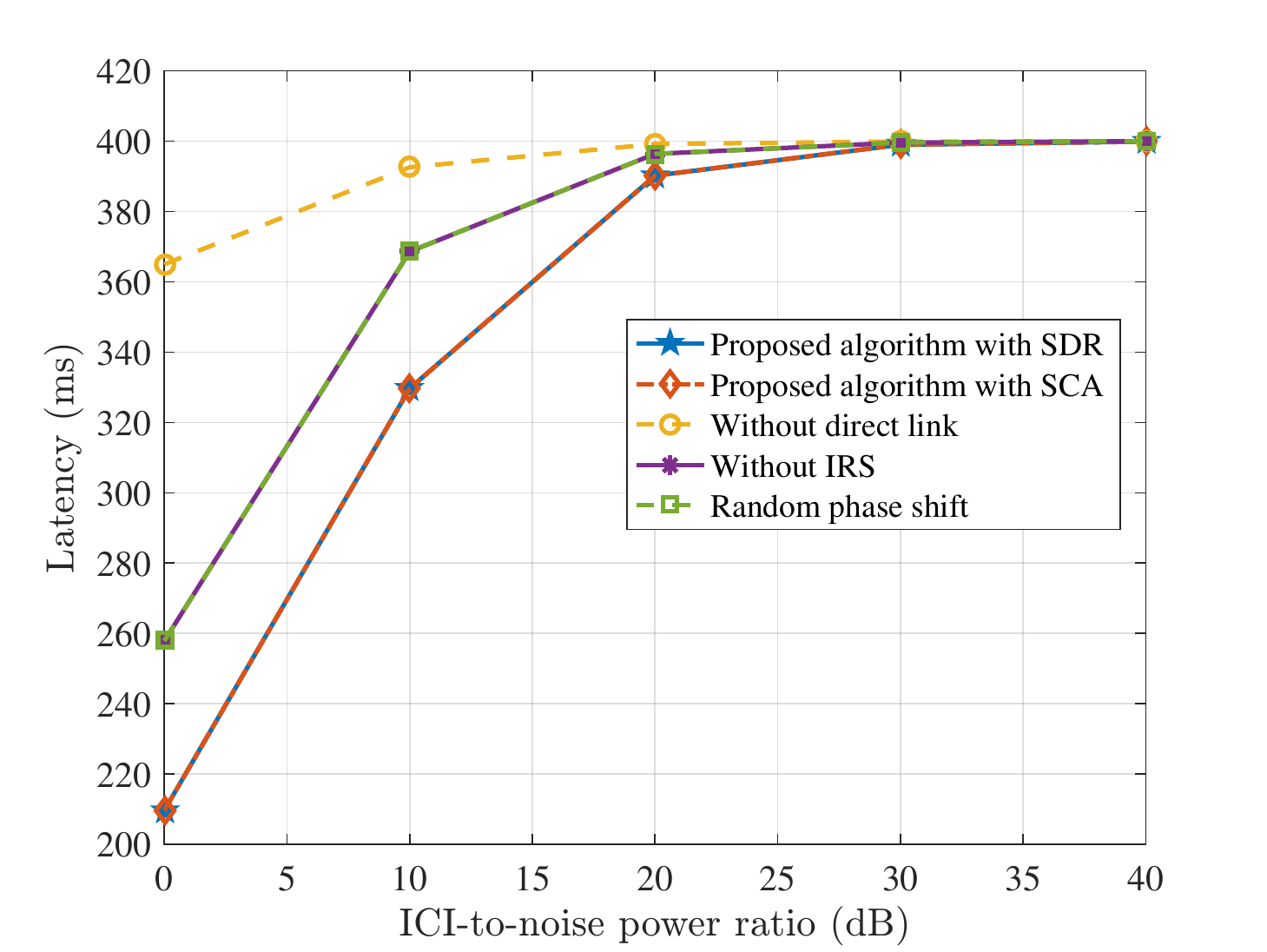}
		\caption{Maximum latency against the ICI-to-noise ratio.}
	\end{minipage}
\end{figure*}

\emph{3) Convergence:} 
To show the convergency of the proposed algorithms, we present the latency against the number of iterations $ I_{o} $ in Fig. 6. It is observed that all the considered algorithms enjoy a fast convergence,  which verifies its practical implementation.

\emph{4) Impact of WDs' Transmit Power:} Fig. 7  shows the latency versus WDs' transmit power $ p_{t} $.  It can be readily observed that, for all schemes, the increase of transmit power $ p_{t} $ leads to a decrease in latency. Moreover, the reduced latency by employing IRSs is significant when WDs' transmit power is moderate. To elaborate, when WDs' transmit power is low (i.e. -20 dBm), the signals reflected by IRSs are so weak that the contribution of employing IRSs is limited. In addition, the reduced latency by employing IRSs is also negligible when the transmit power is high (i.e., 20 dBm), and this is because the WD-BS link is dominant for high transmit power. Additionally, it can be observed that as the transmit power increases, the latency gap between “Without direct link” scheme and “Without IRS” scheme becomes large. Here, “Without direct link” scheme means that BSs only receive signal transmitted from the WD-IRS-BS links, while “Without IRS” means that BSs only receive signals transmitted from the WD-BS links \cite{ref41}. This implies that, compared to “Without direct link” scheme, the “Without IRS” scheme can achieve a lower latency. In this case, the BSs will tend to design sophisticated MUD matrix to receive more power from the WD-BS links, which weakens the role of IRSs.

\emph{5) Impact of Inter-Cell Interference:}  It is worth pointing out that the inter-cell interference (ICI) also affects computing offloading in realistic scenarios. Fig. 8 shows the latency versus ICI-to-noise power ratio. It is observed that, when the ICI-to-noise power ratio is small, upon increasing ICI-to-noise power ratio, the latency increases. Whereas, when the ICI-to-noise power ratio reaches a certain threshold, (i.e., 30 dB), the latency remains unchanged. This is because that, increasing the ICI-to-noise power ratio, WDs will choose to offload less data for edge computing, and when the ICI-to-noise power ratio reaches a certain threshold, all data computing will be performed locally. This demonstrates the necessity to eliminate inter-cell interference.

\section{Conclusion}
In this paper, we aimed to minimize the maximum WD's latency by jointly optimizing offloading data size $ \boldsymbol{\ell} $, edge computing resource $ \mathbf{f}^{e} $, reflecting beamforming vector $ \boldsymbol{\theta} $, and MUD matrix $ \mathbf{W} $, subject to the edge computing capability constraint and IRS phase shift constraint. To address this non-convex problem, we developed a BCD-based algorithm, in which the optimization variables related to the computing and communication setting were solved via an alternating iterative manner. Extensive simulation results verified the benefits of deploying IRSs in MEC systems. In particular, compared with the conventional MEC system, the latency can be effectively reduced from $ 160 $ ms to $ 100 $ ms for distances $ L=60 $ m and $ 100 $ m. Moreover, our simulation results confirmed that our proposed algorithms enjoy a fast convergence, which validates its engineering viability. 

\begin{appendices} 
\section{THE PROOF OF PROPOSITION 1}
For notational convenience, let $\hat{\ell}_{k} \in\left[0, L_{k}\right]$ denote the relaxation of integer value $\ell_{k} \in\left\{0,1, \ldots, L_{k}\right\}$ \cite{ref43}. In addition, fixing $ \mathbf{f}^{e} $, we denote the overall latency of the $ k $-th WD by $\hat{D}_{k}\left(\hat{\ell}_{k}\right) \triangleq \max \left\{D_{k}^{l}\left(\hat{\ell}_{k}\right), D_{k}^{e}\left(\hat{\ell}_{k}\right)\right\}$. Based on (\ref{Eq6}), $ \hat{D}_{k}\left(\hat{\ell}_{k}\right) $ can be re-expressed as
\begin{equation}\label{eq57}
	\hat{D}_{k}\left(\hat{\ell}_{k}\right)= \begin{cases}\frac{\left(L_{k}-\hat{\ell}_{k}\right) c_{k}}{f_{k}^{l}}, & 0 \leq \hat{\ell}_{k} \leq \frac{L_{k} c_{k} R_{k} f_{k}^{e}}{f_{k}^{e} f_{k}^{l}+c_{k} R_{k}\left(f_{k}^{l}+f_{k}^{e}\right)}, \\ \frac{\hat{\ell}_{k}}{R_{k}}+\frac{\hat{\ell}_{k} c_{k}}{f_{k}^{e}}, & \frac{L_{k} c_{k} R_{k} f_{k}^{e}}{f_{k}^{e} f_{k}^{l}+c_{k} R_{k}\left(f_{k}^{l}+f_{k}^{e}\right)}<\hat{\ell}_{k} \leq L_{k}.\end{cases}
\end{equation}
It is obvious that, when $ \hat{\ell}_{k} $ increases from $ 0 $ to $ \frac{L_{k} c_{k} R_{k} f_{k}^{e}}{f_{k}^{e} f_{k}^{l}+c_{k} R_{k}\left(f_{k}^{l}+f_{k}^{e}\right)} $, latency $ \hat{D}_{k}\left(\hat{\ell}_{k}\right) $ reduces, whereas, when $ \hat{\ell}_{k} $ increases from $ \frac{L_{k} c_{k} R_{k} f_{k}^{e}}{f_{k}^{e} f_{k}^{l}+c_{k} R_{k}\left(f_{k}^{l}+f_{k}^{e}\right)} $ to $ L_{k} $, latency $ \hat{D}_{k}\left(\hat{\ell}_{k}\right) $ increases. Thus, when we set  $ \hat{\ell}_{k}=\hat{\ell}_{k}^{*}=\frac{L_{k} c_{k} R_{k} f_{k}^{e}}{f_{k}^{e} f_{k}^{l}+c_{k} R_{k}\left(f_{k}^{l}+f_{k}^{e}\right)} $, the minimum latency of the $ k $-th WD is achieved. Note that the value of offloading data size has to be a non-negative integer, and thus, we can obtain the optimal $ \ell_{k} $ after carrying out the operation  $ \ell_{k}^{*}=\underset{\hat{\ell} \in\left\{\left\lfloor\hat{\ell}_{k}^{*}\right\rfloor,\left[\hat{\ell}_{k}^{*}\right]\right\}}{\arg \min } D_{k}\left(\hat{\ell}_{k}\right) $. 


\section{THE PROOF OF PROPOSITION 2}
We define the value of OF in $ \mathcal{P} 3 $ based on a set of solutions $ (\mathbf{W} , \boldsymbol{\theta}) $ as $ t=g\left( \mathbf{W}, \boldsymbol{\theta}\right)={\max}_{k \in \mathcal{K}}  D_{k}^{e}\left(\mathbf{W}, \boldsymbol{\theta}\right)   $. At the $ \left( l_{2}\right)  $-th iteration, if  $ \mathcal{P} 3.4 $ is feasible, the solution $ \left( \mathbf{W}^{\left( l_{2}\right) } , \boldsymbol{\theta}^{\left( l_{2}+1\right) }\right)  $ is also a feasible solution to problem $ \mathcal{P} 3.1 $. Denote $ \left( \mathbf{W}^{ \left( l_{2} \right) } , \boldsymbol{\theta}^{ \left( l_{2}\right)  }\right)  $ and  $ \left( \mathbf{W}^{\left( l_{2}+1\right) } , \boldsymbol{\theta}^{\left( l_{2}+1\right) }\right)  $ as the optimal solutions to $ \mathcal{P} 3.1 $ at the $\left(  l_{2} \right) $-th and $ \left( l_{2}+1 \right)  $-th iteration, respectively. Since $ \mathbf{W}^{\left( l_{2}+1\right) } $ is the optimal solution of $ \mathcal{P} 3.1 $, then, we have $ g\left(\mathbf{W}^{\left( l_{2}+1\right) }, \boldsymbol{\theta}^{\left( l_{2}+1\right) } \right) \leq g\left(\mathbf{W}^{ l_{2} }, \boldsymbol{\theta}^{ \left( l_{2}+1\right)  } \right) $. As a result, we have the following inequality $ g\left(\mathbf{W}^{\left( l_{2}+1\right) }, \boldsymbol{\theta}^{\left( l_{2}+1\right) } \right) \leq g\left(\mathbf{W}^{ l_{2} }, \boldsymbol{\theta}^{\left( l_{2}+1\right) } \right) \leq \left(\mathbf{W}^{ l_{2} }, \boldsymbol{\theta}^{ l_{2} } \right) $.

\end{appendices}

\end{document}